\documentclass[epj,nopacs,fleqn]{svjour}
\pdfoutput=1
\usepackage{graphicx,amsmath,amssymb,slashed,cite,verbatim,url}
\usepackage{color}

\newcommand{\nn}{\nonumber}
\newcommand{\perc}{\%}
\newcommand\sss{\scriptscriptstyle}

\begin{document}

%%%%%%%%%%%% Begin Cover Page %%%%%%%%%%%%%%%%%%%%%%%%%%%%%%%%%%%%%%%%%%%%
\title{Higgs production in association with a single top quark~at~the~LHC}

\author{
 Federico Demartin\inst{1},
 Fabio Maltoni\inst{1},
 Kentarou Mawatari\inst{2},
 Marco Zaro\inst{3,4}
}

\institute{ 
 Centre for Cosmology, Particle Physics and Phenomenology (CP3),
 Universit\'e catholique de Louvain,\\
 B-1348 Louvain-la-Neuve, Belgium
 \and
 Theoretische Natuurkunde and IIHE/ELEM, Vrije Universiteit Brussel,
 and International Solvay Institutes,\\
 Pleinlaan 2, B-1050 Brussels, Belgium
 \and
 Sorbonne Universit\'es, UPMC Univ. Paris 06, UMR 7589, LPTHE, 
 F-75005, Paris, France
 \and
 CNRS, UMR 7589, LPTHE, F-75005, Paris, France
}

\abstract{
We present a detailed study of Higgs boson production in association
with a single top quark at the LHC, at next-to-leading order accuracy in
QCD. 
We consider total and differential cross sections, at the parton level
as well as by matching short distance events to parton showers, for both
$t$-channel and $s$-channel production.  
We provide predictions relevant for the LHC at 13 TeV together with a
thorough evaluation of the residual uncertainties coming from scale
variation, parton distributions, strong coupling constant and heavy
quark masses. 
In addition, for $t$-channel production, we compare results as obtained
in the 4-flavour and 5-flavour schemes, pinning down the most relevant
differences between them. 
Finally, we study the sensitivity to a non-standard-model relative phase
between the Higgs couplings to the top quark and to the weak bosons. 
}  

\date{}

\titlerunning{Higgs production in association with a single top quark at
the LHC}   
\authorrunning{F.~Demartin et al.}

\maketitle

%%%%%%%%%%%% End of Cover Page %%%%%%%%%%%%%%%%%%%%%%%%%%%%%%%%%%%%%%%%%

\vspace*{-12cm}
\noindent
\small{MCnet-15-07, CP3-15-08}
\vspace*{10.4cm}

%%%%%%%%%%%%%% Begin Section: Intro %%%%%%%%%%%%%%%%%%%%%%%%%%%%%%%%%%%%%%%%% 
\section{Introduction}\label{sec:intro}

The first Run of the LHC has already collected compelling evidence that
the scalar particle observed at 125 GeV is the one predicted by the
Brout--Englert--Higgs symmetry breaking
mechanism~\cite{Englert:1964et,Higgs:1964pj} of $SU(2)_L \times U(1)_Y$
as implemented in the Standard Model (SM)~\cite{Weinberg:1975gm}.  
In such minimal case, the strengths of the Higgs boson couplings to the
elementary particles, including the Higgs boson itself, are uniquely
determined by their masses. 
While somewhat limited and subject to additional ad hoc assumptions, the
first measurements of the Higgs couplings to fermions and vector bosons
agree well with the SM predictions~\cite{cms2013,atlas2013}. 

Such general agreement with the SM expectations {\it and} the absence of
any evidence (from the LHC itself) of the existence of new states at
the TeV scale, motivate a thorough study of the Higgs boson
interactions at the Run II.  
In addition to the coupling strength determinations conducted so far, the Lorentz
structure of the vertices as well as the possible existence of a
relative phase among the couplings need to be fully assessed.
In order to gather the necessary information, the widest possible
campaign of measurements has to be undertaken, including different
production and decay modes of the Higgs boson.
In addition, given the limited discriminating power of single channels, 
a global combination of the relevant measurements will be necessary. 
To achieve this goal at the LHC one needs to adopt a complete and
consistent theoretical framework, able to encompass interactions that go
beyond the SM (and possibly to organise them in terms of an ordering
principle), and that allows the systematic inclusion of higher-order
corrections, both QCD and electroweak (EW).  
This latter point is a conditio-sine-qua-non at the LHC,
in order to control total rates and
differential distributions and to estimate the residual uncertainties. 
Such a theoretical framework exists and amounts to ``simply'' extend 
the dimension-4 SM Lagrangian to all operators of higher dimensions (up 
to dimension-6 in this first instance) consistent with the unbroken SM
symmetries $SU(3)_C \times SU(2)_L \times U(1)_Y$; {\it i.e.} to
consider the SM as an effective field theory valid up to a scale
$\Lambda$~\cite{Buchmuller:1985jz,Grzadkowski:2010es}. 

\begin{figure*}
\center 
\includegraphics[height=0.11\textwidth]{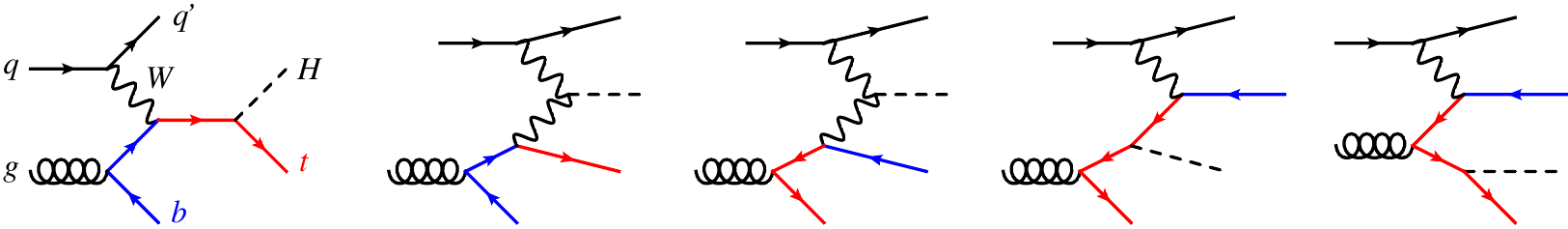}\\[2mm]
\includegraphics[height=0.1\textwidth]{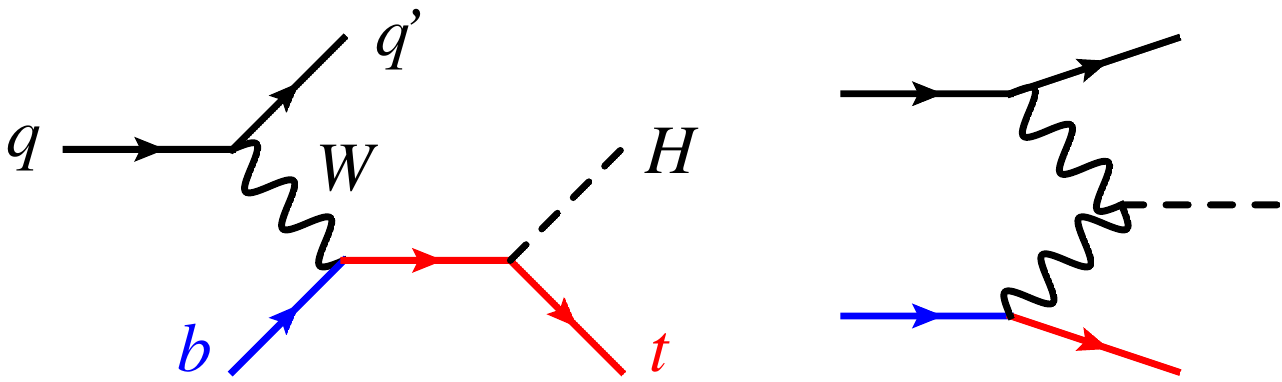}
\caption{LO Feynman diagrams for $t$-channel $tH$ production in the 4F
 scheme (top) and in the 5F scheme  (bottom).} 
\label{fig:diagram-t}
\end{figure*}

\begin{figure}
\center 
\includegraphics[height=0.1\textwidth]{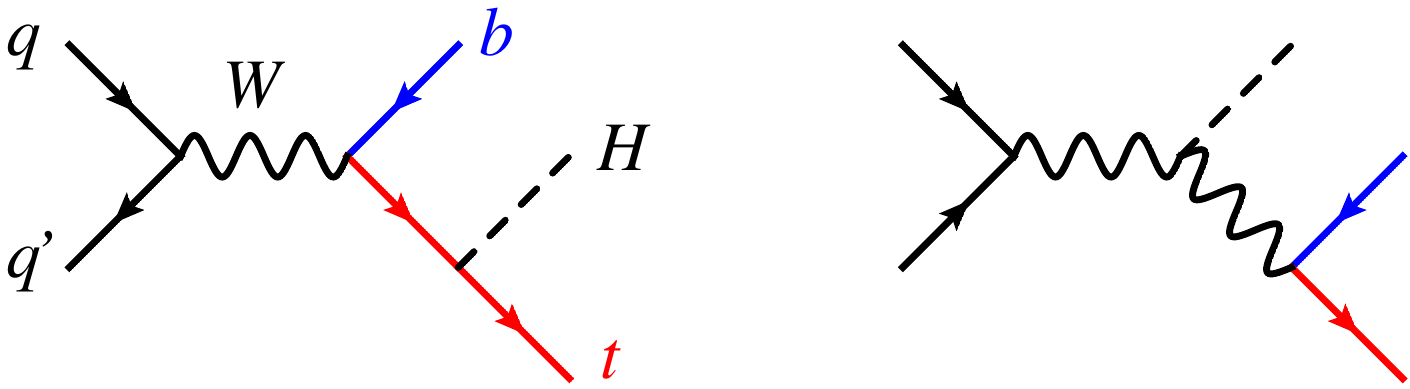}
\caption{LO Feynman diagrams for $s$-channel $tH$ production.}  
\label{fig:diagram-s}
\end{figure}

This work fits in the above general strategy and focuses on Higgs
production in association with a single top quark. 
As in single top production, at the leading order (LO) in QCD
one can organise the production mechanisms into three
groups, based on the virtuality of the $W$ boson: $t$-channel production
(fig.~\ref{fig:diagram-t}), $s$-channel production
(fig.~\ref{fig:diagram-s}), and associated production with an on-shell
$W$ boson.  
While characterised by a rather small cross section with respect to the
main single Higgs production channels (gluon--gluon fusion, vector boson
fusion and associated production, and $t\bar t H$), Higgs and single-top
associated production features unique aspects that make this process
particularly interesting for Higgs
characterisation~\cite{Biswas:2012bd,Farina:2012xp}.
Notably, it is among the very few processes relevant for LHC
phenomenology (together with $H\to \gamma \gamma$ and 
$gg \to ZH$) to be sensitive to the relative size and phase of the
coupling of the Higgs boson to the top quark and to the weak bosons. 
For $t$-channel and $W$-boson associated production, diagrams where the
Higgs couples to the top quark interfere destructively with those where
the Higgs couples to the $W$ boson (due to the unitarity of the weak
interactions in the SM), making cross sections and distributions
extremely sensitive to departures of the Higgs couplings from the SM
predictions~\cite{Maltoni:2001hu}.  

The aim of the first part of this work is to provide accurate SM
predictions including  QCD corrections at next-to-leading order (NLO)
for $t$- and $s$-channel Higgs production in association with a single
top quark, as well as reliable estimates for the residual uncertainties
in rates and distributions.  
Particular attention is devoted to the uncertainty related to the
different flavour schemes that can be adopted to compute
the dominant $t$-channel production mode. 
The corresponding SM predictions are the necessary theoretical input to
possibly assess the existence of deviations due to new physics 
(be it resonant or not);
to this aim, the study of the uncertainties in total rates as well as in
differential distributions becomes of primary importance. 

We then consider how accurately and precisely the effects of the (only)
dimension-6 operator that modifies the value and the phase of the top quark
Yukawa coupling can be predicted, again at the total as well as at the
differential level. 
This information is useful to assess the reach of the LHC to constrain
the relevance of this dimension-6 operator ({\it i.e.} to bound the complex coefficient in front)
and, if deviations from the SM are detected, to quantify them. 

The paper is organised as follows. 
In sect.~\ref{sec:tH} we introduce the main features of the Higgs and top quark
associated production.
In sect.~\ref{sec:tchannel} we focus on the $t$-channel production mode, with a
special attention to the issues connected to the 4-flavour (4F) and
5-flavour (5F) schemes.
We describe the settings of the calculation, 
present results in the SM for total rates up to NLO in QCD and their uncertainties, 
and finally show relevant differential distributions at NLO
matched to a parton shower. 
In sect.~\ref{sec:schannel} we shortly consider the $s$-channel production
mechanism, which has a much smaller impact on Higgs phenomenology in the SM.
We evaluate the total cross sections and its uncertainties in the SM and
show some representative distributions in comparison with the
corresponding $t$-channel ones. 
In sect.~\ref{sec:HC} we study the impact of an anomalous,
CP-violating top quark Yukawa interaction on $t$-channel production,
both at the total and differential cross section level. 
We summarise our findings in sect.~\ref{sec:summary}.

%%%%%%%%%%%%%% Begin Section: tH production %%%%%%%%%%%%%%%%%%%%%%%%%%%%
\section{Main features}\label{sec:tH}

In this section we introduce the main features of Higgs production in association with a single top quark.
As already mentioned in the introduction, at LO in QCD one can effectively organise the various 
production mechanisms into three groups, based on the virtuality of the $W$ boson: 
$t$-channel production features a space-like $W$, 
$s$-channel production a time-like $W$, 
and $W$-associated production an on-shell $W$ boson. 
One has to bear in mind that while this classification is certainly useful, 
it is not physical, being  an approximation that holds only at LO and in the 
5-flavour scheme. 
At higher orders in QCD, or using a different flavour scheme to define the processes, 
the separation becomes increasingly fuzzy,
as it will be clarified at the end of this section.

As in single top production in the SM, 
$tH$ production is always mediated by a $tWb$ vertex and therefore it
entails the presence of a bottom quark either in the initial 
($t$-channel and $W$-associated) or in the final state ($s$-channel).
In the case of initial-state bottom quarks, two different approaches, 
the so-called 4F and 5F schemes, 
can be followed to perform perturbative calculations. 

In the 4F scheme one assumes that the typical scale of the hard process ${\cal Q}$ 
is not significantly higher than bottom quark mass, 
which in turn is considerably heavier than $\Lambda_{\rm QCD}$,
${\cal Q} \gtrsim m_b \gg \Lambda_{\rm QCD}$.
Technically, one constructs an effective theory of QCD with only four light flavours, 
where heavier quarks (bottom and top), being massive, do not contribute
to the initial-state proton wave-function (in terms of parton
distribution functions (PDFs)), 
nor to the running of the strong coupling, and they appear only as final-state particles. 
In so doing, mass effects in the kinematics of heavy-quark production are correctly taken into account 
already at the lowest order in perturbation theory. 
In addition, the matching to parton-shower programs is straightforward, 
the heavy-quark mass acting as an infrared cutoff for inclusive observables. 
However, limitations might arise when ${\cal Q}\gg m_b$ and one probes kinematic configurations 
which are dominated by almost collinear  $g \to b \bar b$  splittings: 
in this case the accuracy of predictions can be spoiled by large logarithms $\log ({\cal Q}^2/m_b^2)$ 
appearing at all orders in perturbative QCD. 
Were this the case, such large logarithms would harm the behaviour of a fixed order expansion in $\alpha_s$.

This issue can be addressed in the 5F scheme (and improvements thereof), 
whose aim is to reorganise the perturbative expansion by resumming such logarithms
via the DGLAP equations. 
One starts by assuming ${\cal Q}\gg m_b$ and defines a scheme where power corrections 
of order $m_b/{\cal Q}$ appear at higher orders in the $\alpha_s$ expansion. 
In practice, one sets the bottom mass to zero and includes bottom quarks in the initial state as 
proton constituents.\footnote{%
The bottom mass can be reinstated explicitly at higher-orders by systematically including it in diagrams that do not feature bottom quarks in the initial state, the so-called S-ACOT scheme~\cite{Kramer:2000hn}. In this work we adopt a ``pure'' 5F scheme where  $m_b=0$ throughout.}
In so doing, towers of logarithms associated with the initial-state $g \to b \bar b$ splitting 
are resummed to all orders in perturbation theory by evolving the
perturbative bottom quark PDF 
via the DGLAP equations. 

Computations in the 5F scheme are typically much simpler than the corresponding 4F ones, 
because of the lesser final-state multiplicity and the simpler phase space. 
This is for example the reason why single-top production is known at NNLO in the 5F scheme~\cite{Brucherseifer:2014ama} while only at NLO in the 4F~\cite{Campbell:2009ss}. 
For a systematic investigation of the sources of differences between the 4F and 5F schemes 
in single $b$-quark and double $b$-quark induced processes we refer the reader to~\cite{Maltoni:2012pa} 
and~\cite{Wiesemann:2014ioa}, respectively. 
In short, the 4F and 5F schemes differ in what kind of terms are pushed into the missing higher-order corrections. 
Therefore, as the accuracy of the predictions for a given observable increases, milder differences 
should be expected between the schemes. This provides a strong motivation to go at least to NLO accuracy 
in the computation of the $t$-channel cross section, in order to reduce the flavour-scheme dependence 
of the predictions and thus the overall theoretical uncertainty. 
The final accuracy, however, will depend on the specific observable considered, 
whose perturbative accuracy can be different in the two schemes.

In the case of (Higgs and) single top production at hadron colliders, 
the 5F scheme has also the operational advantage that allows an easy separation of the various
production mechanisms into the three groups mentioned above. 
In the 5F scheme the $t$-channel, $s$-channel and $W$-associated production are independent up to NLO and start to interfere only at NNLO, and the $W$-associated production interferes with $t \bar t H$ starting from NLO. 
In the 4F, on the other hand, the $t$-channel at NLO can interfere with the $s$-channel (at NNLO)
and with $W$-associated production (if the $W$ decays hadronically), and
the $W$-associated production also interferes with $t \bar t H$ already at the tree level.
While the former interferences are very small and can be safely neglected if the aim is to evaluate 
the dominant $t$-channel cross section, 
the interference of $W$-associated production with $t \bar t H$ turns out instead to be quite large. 
The on-shell $W$ associated production therefore needs a dedicated study that we defer to a separate work.

%%%%%%%%%%%%%%%%%%%%%%%%%%%%%%%%%%%%%%%%%%%%%%%%%%%%%%%%%%%%
\section{$\boldsymbol{t}$-channel production}
\label{sec:tchannel}

In this section we present the SM predictions for $t$-channel Higgs plus
single top production at the LHC (see fig.~\ref{fig:diagram-t}), at NLO
accuracy in QCD. 
We first describe the technical setup we have used for NLO simulations,
the input parameters as well as the various sources of theoretical
uncertainties.
We then show results for the inclusive cross section at the LHC with
$\sqrt{s}=13$~TeV, discussing how to combine the theoretical
uncertainties, and finally present NLO distributions matched to
parton shower.

%%%%%%%%%%%%%% Begin subsection: NLO simulations %%%%%%%%%%%%%%%%%%%%%% 
\subsection{NLO simulations, parameters and uncertainties}
\label{sec:nlosetup}

In this work, we employ the {\sc MadGraph5\_aMC@NLO}
framework~\cite{Alwall:2014hca}, which allows to compute both inclusive
cross sections and differential distributions matched to parton-shower
programs, up to NLO accuracy in QCD, in a fully automatic
way~\cite{Frixione:2002ik,Ossola:2007ax,Frederix:2009yq,Hirschi:2011pa} once the
relevant Feynman rules and UV/$R_2$ counterterms for a given theory are
provided in the form of a UFO
model~\cite{Degrande:2011ua,Alloul:2013bka,deAquino:2011ub}. While these
extra Feynman rules are available in {\sc MadGraph5\_aMC@NLO} by default
for the SM, non-SM interactions that will be considered later in
sect.~\ref{sec:HC} are encoded in the {\sc HC\_NLO\_X0}
model~\cite{Artoisenet:2013puc,Maltoni:2013sma,Demartin:2014fia},
publicly available online in the {\sc FeynRules} repository~\cite{FR-HC:Online}. 

In {\sc MadGraph5\_aMC@NLO}
the code and events for $t$-channel $tH$ production at hadron colliders
in the 4F scheme can be automatically generated by issuing the following
commands:
\begin{verbatim}
(> import model loop_sm)
 > generate p p > h t b~ j $$ w+ w- [QCD]
 > add process p p > h t~ b j $$ w+ w- [QCD]
 > output
 > launch
\end{verbatim}
while the corresponding commands in the 5F scheme are:
\begin{verbatim}
 > import model loop_sm-no_b_mass
 > define p = p b b~
 > define j = p
 > generate p p > h t j $$ w+ w- [QCD]
 > add process p p > h t~ j $$ w+ w- [QCD]
 > output
 > launch
\end{verbatim}
Note that the \texttt{\$\$ w+ w-} syntax removes $s$-channel $tH$
diagrams as well as real-correction diagrams where an on-shell $W$ decays
to two light quarks, which belong to $W$-associated production. 
The top quark decays are subsequently performed starting from the event file 
(in the Les Houches format~\cite{Alwall:2006yp}) by
{\sc MadSpin}~\cite{Artoisenet:2012st}, following a
procedure~\cite{Frixione:2007zp} that keeps spin correlations.  

In the numerical calculation, the mass of the Higgs boson is set to
$m_{H}=125.0$~GeV, while the mass of the top quark is set to
$m_{t}=173.3$~GeV.  
We renormalise the top quark Yukawa coupling on-shell, setting it to
$y_t/\sqrt{2}=m_t/v$, 
where $v\sim246$~GeV is the EW vacuum expectation value.

The on-shell mass of the bottom quark is set to 
\begin{align}
m_{b} = 4.75 \pm 0.25~\mathrm{GeV} \,,
\label{eq:mb}
\end{align}
where we take the uncertainty to be of ${\cal O}(\Lambda_{\rm QCD})$,
accordingly to the prescription in ref.~\cite{Martin:2010db}.
On the other hand, we set the bottom quark Yukawa coupling to zero,
because effects related to the $Hb\bar b$ interactions are negligible for this process.
We remind that in the 4F scheme the value of $m_b$ enters the
hard-scattering matrix element and the final-state phase space, while in
the 5F scheme it affects only the parton luminosity.

PDFs are evaluated by using three global
fits: NNPDF2.3~\cite{Ball:2012cx}, MSTW2008~\cite{Martin:2009iq} and
CT10~\cite{Lai:2010vv}, through the LHAPDF
interface~\cite{Whalley:2005nh}.  
PDF uncertainties are computed for each PDF set, following the 
recipes summarised in~\cite{Alekhin:2011sk}.  
A comparison among these three global fits allows to estimate the PDF
systematic uncertainties related to the technical details of the fitting
procedure employed by each group. 
We note that the above three PDF collaborations provide NLO PDF sets both in the 4F and 5F
schemes, while only MSTW gives LO PDFs in both the schemes.

\begin{figure*}
\center 
 \includegraphics[width=1\columnwidth]{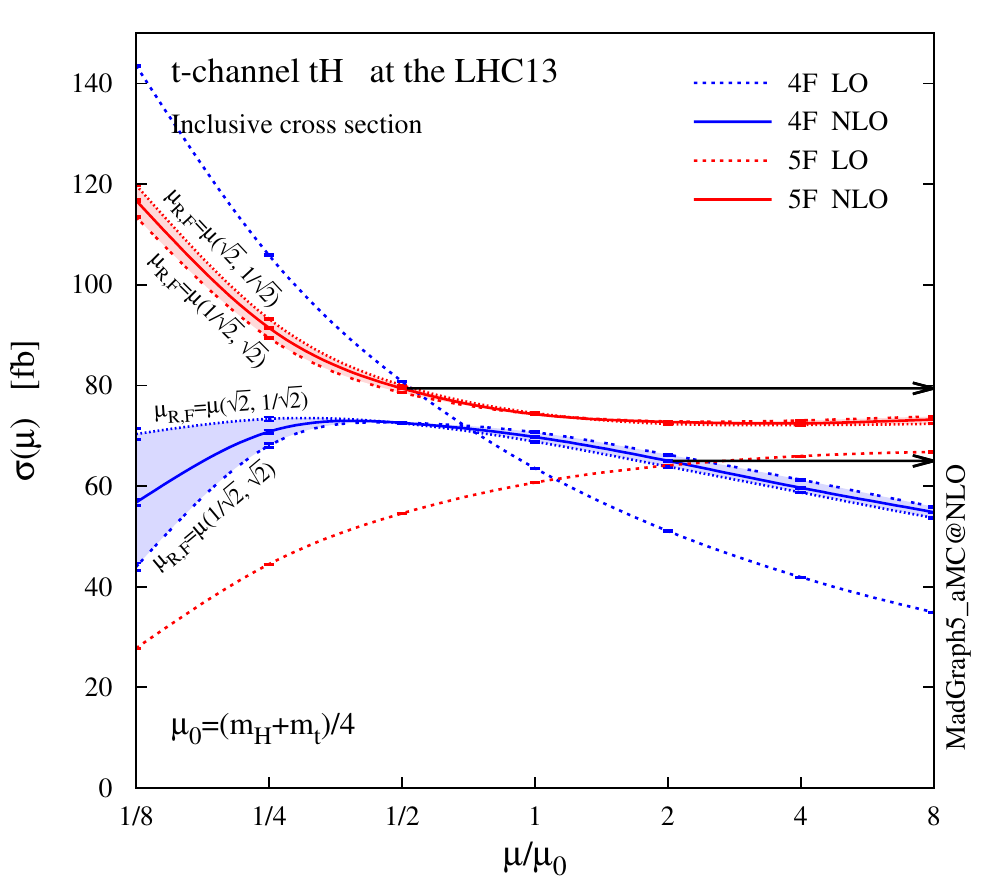}
 \includegraphics[width=1\columnwidth]{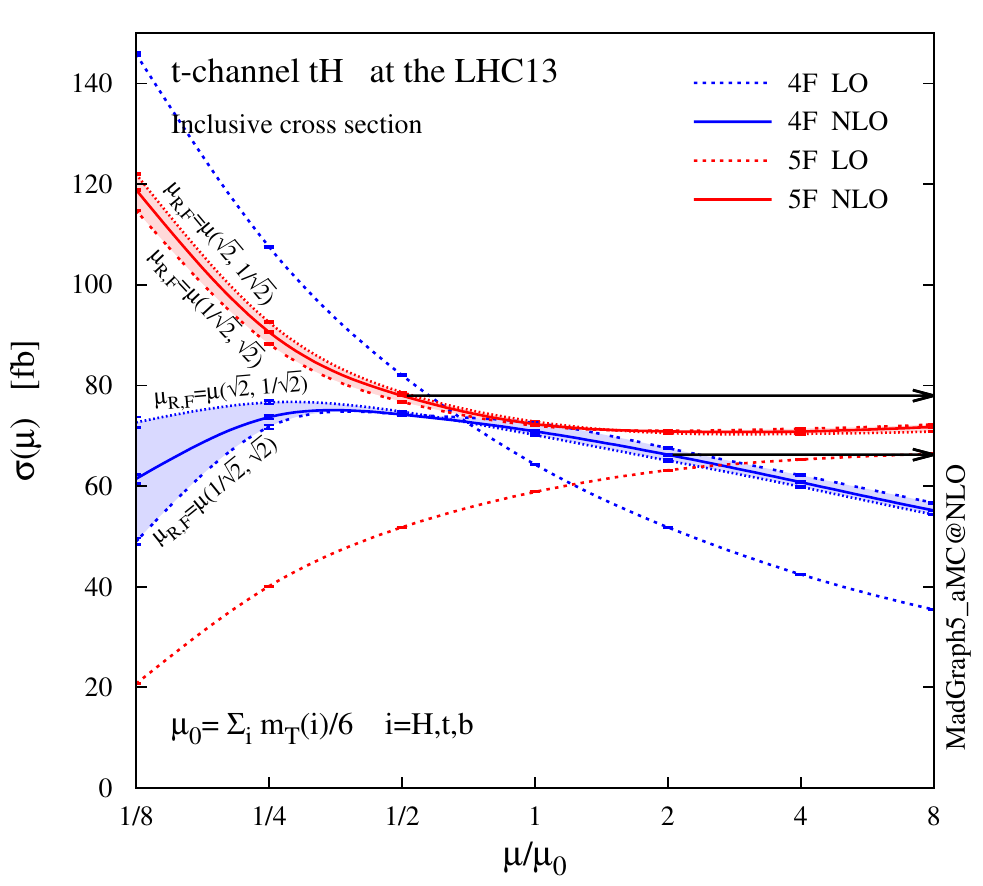} 
 \caption{Scale dependence of the total cross sections for the
 $pp \to tHq + \bar tHq$ production at the 13-TeV LHC, 
 where the 4F (blue) and  5F (red) schemes are compared.
 LO (dashed) and NLO (solid) predictions with MSTW2008 LO/NLO PDFs are
 presented for $\mu_R=\mu_F \equiv \mu \,$, with a static (left figure) and a
 dynamic (right figure) scale choice.
 Two off-diagonal profiles of the scale dependence at NLO are also shown, 
 for $( \mu_R=\sqrt{2}\mu\,,\,  \mu_F=\mu/\sqrt{2} )$ and for
 $( \mu_R=\mu/\sqrt{2}\,,\,  \mu_F=\sqrt{2}\mu ) \,$. 
 The black arrows visualise the envelope of the combined scale and
 flavour-scheme uncertainty defined in
 eq.~\eqref{eq:enevelope_MU-FS}.}  
\label{fig:scaledep}
\end{figure*} 

The reference value for the strong coupling constant we employ here is
\begin{align}
 \alpha_s^{(\rm{NLO})}(m_Z) = 0.1190 \pm 0.0012\,,
\label{eq:asnlo5F}
\end{align}
where the uncertainty is taken accordingly to the PDF4LHC
recommendation~\cite{Alekhin:2011sk,Botje:2011sn}, and the central value
is chosen such that our $68 \perc$ confidence interval encompasses the
current PDG world average~\cite{Agashe:2014kda} and the best
$\alpha_s(m_Z)$ estimates obtained by each of the three PDF global
fits~\cite{Martin:2009bu,Lai:2010nw,Lionetti:2011pw}. 
We remark that the value in eq.~\eqref{eq:asnlo5F} is consistent with
the 5F description.  
Since the difference between 4F and 5F in the $\alpha_s$ running is
limited to scales above $m_b$, eq.~\eqref{eq:asnlo5F} can be translated
into the following condition on $\alpha_s(m_b)$ (running $\alpha_s$ at
2-loop accuracy) 
\begin{align}
 \alpha_s^{(\rm{NLO})}(m_b) = 0.2189 \pm 0.0042\,,
\label{eq:asMBnlo}
\end{align}
which is now flavour-scheme independent.

CT10 does not provide PDF sets to compute $m_b$ uncertainties in the 5F
scheme and PDF uncertainties in the 4F scheme; 
both CT10 and MSTW2008 do not provide 4F PDF sets with different
$\alpha_s(m_Z)$ values.
Thus, it is possible to address all the various sources of uncertainty
in both schemes only when using NNPDF2.3 parton distributions, while
MSTW2008 and CT10 uncertainty bands can be sometimes underestimated
(though just slightly, as we will see later in sect.~\ref{sec:totrates}).  

For matching short-distance events to parton shower we use the {\sc MC@NLO} method~\cite{Frixione:2002ik} 
with  {\sc Pythia8}~\cite{Sjostrand:2007gs}, while 
{\sc HERWIG6}~\cite{Corcella:2000bw} has been used for a few comparisons. 
We recall that matching to {\sc Pythia6}~\cite{Sjostrand:2006za} (virtuality-ordered, or
$p_T$-ordered for processes with no final-state radiation) and 
{\sc HERWIG++}\cite{Bahr:2008pv} are also available inside {\sc MadGraph5\_aMC@NLO}.
Jets are reconstructed by means of the anti-$k_T$
algorithm~\cite{Cacciari:2008gp} as implemented in
{\sc FastJet}~\cite{Cacciari:2011ma}, with distance parameter $R=0.4$,  
and required to have
\begin{align}
 p_T(j)>30~{\rm GeV}\,, \quad |\eta(j)|<4.5\,.
\label{eq:jet_definition}
\end{align}
A jet is identified as $b$-jet if a $b$-hadron (or $b$-quark for
fixed-order calculations) is found among its constituents, and if the jet satisfies
\begin{align}
 p_T(j_b)>30~{\rm GeV}\,, \quad |\eta(j_b)|<2.5\,.
\label{eq:bjet_definition}
\end{align}
We assume 100\% $b$-tagging efficiency in this work.

%%%%%%%%%%%%%% Begin subsection: Total rates %%%%%%%%%%%%%%%%%%%%%%%%%% 
\subsection{Total rates}\label{sec:totrates}

In this section we present the total cross section for $t$-channel
production of a Higgs boson together with a single top quark (or
antiquark), at NLO in QCD.
The main sources of theoretical uncertainty that we address here are:
\begin{itemize}
 \item renormalisation and factorisation scale dependence,
 \item 4F and 5F scheme dependence,
 \item PDF uncertainty,
 \item $\alpha_s(m_Z)$ uncertainty,
 \item $m_b$ uncertainty.
\end{itemize}
At the end of this section we will also briefly comment on the impact of the
bottom quark Yukawa coupling and of the dependence of the results 
on the Higgs and the top quark masses. 

We start by showing in fig.~\ref{fig:scaledep} the renormalisation and factorisation scale
dependence of the LO and NLO total cross sections, both in the 4F and
5F schemes. 
We compute cross sections with two different scale choices, and vary
$\mu_R=\mu_F \equiv \mu$ around a central scale $\mu_0$ which is chosen as 
\begin{align}
 \mu_0^{s} = (m_H + m_t)/4
\label{eq:mu0fix}
\end{align}
for the static scale choice (left figure), and 
\begin{align}
 \mu_0^{d} = H_T/6 = \sum_{i=H,t,b} m_T(i)/6
\label{eq:mu0dyn}
\end{align}
for the event-by-event dynamic choice (right figure), where
$m_T\equiv\sqrt{m^2+p_T^2}$ is the transverse mass of a particle. 

\begin{table}
\center
\begin{tabular}{l|llll}
 \hline
    \rule{0pt}{3ex}
 scheme
  & $\sigma_{\rm LO}$~{\small [fb]}
  & $\sigma_{\rm NLO}$~{\small [fb]}
  & $K$
    \\[0.7ex]
\hline
     \rule{0pt}{3ex}
 4F\ ($\mu_0^{s}$)  
     & 63.46(8) $^{+27.2\%}_{-19.7\%}$
     & 69.43(7) $^{+4.0\%}_{-5.8\%}$
     & 1.09
 \\[0.3ex]
     \rule{0pt}{3ex}
 5F\ ($\mu_0^{s}$) 
     & 60.66(6) $^{+5.6\%}_{-10.0\%}$
     & 73.45(8) $^{+7.0\%}_{-2.3\%}$
     & 1.21
 \\[0.7ex]
 \hline
     \rule{0pt}{3ex}
 4F\ ($\mu_0^{d}$)  
     & 64.31(8) $^{+27.6\%}_{-19.5\%}$
     & 71.29(10) $^{+3.8\%}_{-7.1\%}$
     & 1.11
 \\[0.3ex]
     \rule{0pt}{3ex}
 5F\ ($\mu_0^{d}$) 
     & 58.83(5) $^{+7.6\%}_{-11.9\%}$
     & 71.54(7) $^{+7.3\%}_{-2.1\%}$
     & 1.22
 \\[0.7ex]
 \hline
\end{tabular}
 \caption{LO and NLO cross sections and corresponding $K$ factors for 
 $t$-channel $tH$ production at the 13-TeV LHC in the 4F and 5F schemes.
 MSTW2008 PDFs have been used.
 The integration error in the last digit(s) and the scale dependence by a
 factor 2 around the static and dynamic scale choices in
 eqs.~\eqref{eq:mu0fix} and \eqref{eq:mu0dyn} are also reported.}
\label{tab:xsect_LOvsNLO}
\end{table}

We find a pattern similar to the case of the single top production (see
fig. 3 in~\cite{Campbell:2009ss}).
At LO the scale dependence in the 4F scheme is stronger than in the 5F,
simply because the 4F calculation starts already at order $\alpha_s$.
As expected, predictions at NLO are much more stable under the scale
variation than at LO. 
We find that the 4F and 5F predictions are in better agreement
if $\mu$ is chosen to be roughly a factor 4 (6) smaller than the typical
hard scale of the process $m_H+m_t$ ($H_T$) for the static (dynamic)
scale choice.  
This is a known and general feature of $b$-initiated processes at hadron
colliders~\cite{Maltoni:2012pa}.
At such reduced scales the 4F and 5F predictions are typically in good agreement, 
and this is indeed what we observe taking the reference scale choice $\mu_0$ 
as in eqs.~\eqref{eq:mu0fix} and \eqref{eq:mu0dyn}. 
Table~\ref{tab:xsect_LOvsNLO} shows the corresponding values of the LO and NLO 
cross sections in fig.~\ref{fig:scaledep}, where the uncertainty from missing 
higher orders is estimated varying the scale $\mu$ by a factor 2 around $\mu_0$.

In fig.~\ref{fig:scaledep} we also plot two off-diagonal 
($\mu_R\ne\mu_F$) slices of the NLO cross section surface in the plane
($\mu_R,\mu_F$), shifted by a factor $\sqrt{2}$ in the direction orthogonal to the diagonal.
The effects of off-diagonal scale choices are more pronounced in the 4F scheme than in the 5F, 
even though in general they are quite modest, except at very low scales, {\it i.e.} comparable to $m_b$.  
We conclude that, for our choice of $\mu_0$, the diagonal $\mu_R=\mu_F$
is sufficiently representative of the scale dependence of the total
cross section, when the scale is varied by the usual factor two. 
We also observe that the scale value which minimises the flavour-scheme
dependence is rather stable under shifts away from the diagonal.

We note that the scale dependence pattern is strongly correlated to
the flavour scheme employed. Therefore, after we estimate the scale dependence 
of both 4F and 5F results (varying the scale $\mu_F=\mu_R\equiv\mu$ 
by a factor 2 around $\mu_0$), we define a combined scale and flavour-scheme
uncertainty band by taking the envelope of the
extremal points (shown by the black arrows in fig.~\ref{fig:scaledep}),
and the best prediction for the cross section as the central point of
this envelope. 
The total cross section at NLO and its combined scale plus flavour-scheme uncertainty 
are defined by 
\begin{align}
 \sigma_{\rm NLO}  = ( \sigma^{+}+\sigma^{-} )/2 \,, \quad
 \delta_{\rm \mu+FS} = ( \sigma^{+}-\sigma^{-} )/2 \,,
\label{eq:enevelope_MU-FS}
\end{align}
where
\begin{align}
 \sigma^{+} & = \max \limits_{\mu \in [\mu_0/2, \, 2\mu_0]} 
 \big\{\, \sigma^{\rm 4F}_{\rm NLO}(\mu) \,,\, 
          \sigma^{\rm 5F}_{\rm NLO}(\mu) \,\big\} \,, \\
 \sigma^{-} & = \min \limits_{\mu \in [\mu_0/2, \, 2\mu_0]}  
 \big\{ \sigma^{\rm 4F}_{\rm NLO}(\mu) \,,\, 
        \sigma^{\rm 5F}_{\rm NLO}(\mu) \big\} \,.
\end{align}

Now we turn to the PDF, $\alpha_s(m_Z)$ and $m_b$ uncertainties.
In principle these three uncertainties can be correlated.
However, the correlations are very small and can be often neglected in
combinations.
For example, using NNPDF, we have explicitly checked
that the combined PDF+$\alpha_s$ uncertainty computed with
full correlations differs from the one without correlations 
by $0.1\perc$ at most.
In the 4F scheme $m_b$ is independent of PDF and $\alpha_s$, while we
confirmed that the uncertainty correlation between PDF and $m_b$ in the
5F scheme is well below the percent level.
Moreover, the correlation between $\alpha_s$ and $m_b$ is tiny and can
be neglected~\cite{Martin:2010db}. 
We note that neglecting correlations allows us to compare PDF uncertainty bands at a common $\alpha_s$ value,  once central predictions (computed with this common $\alpha_s$) are dressed with their corresponding fractional PDF uncertainty (computed with each group's dedicated set). This is a known fact and it has been extensively used in recent PDF benchmarks~\cite{Ball:2012wy}.

Given that correlations among the uncertainties are very
small, as discussed above, and also that not every PDF set allows to
take into account all the correlations, we define the combined PDF,
$\alpha_s$ and $m_b$ uncertainty by simply summing the uncertainties in 
quadrature as
\begin{align}
\delta_{{\rm PDF}+\alpha_s+m_b}^{\pm} = \sqrt{ \left( \delta_{\rm PDF}^{\pm} \right)^2 + 
\left( \delta_{\alpha_s}^{\pm} \right)^2 + \left( \delta_{m_b}^{\pm} \right)^2} \,.
\label{eq:pdfasmb_totunc}
\end{align}
Finally, we define the total theoretical uncertainty as the linear sum 
of the upper and lower variations for $\delta_\mu$ and $\delta_{{\rm PDF}+\alpha_s+m_b}$ 
in a given flavour scheme. 

\begin{table*}
\center
\begin{tabular}{r|llclll|llclll}
 \hline
    \rule{0pt}{3ex}
    $t$-channel
  & $\sigma_{\rm NLO}^{(\mu_0^s)}$~{\small [fb]} 
  & $\delta^\perc_{\mu}$ \hspace*{-1em}
  & $\delta^\perc_{{\rm PDF}+\alpha_s+m_b}$ \hspace*{-1em}
  & $\delta^\perc_{\rm PDF}$ \hspace*{-1em}
  & $\delta^\perc_{\alpha_s}$ \hspace*{-1em}
  & $\delta^\perc_{m_b}$ 
  & $\sigma_{\rm NLO}^{(\mu_0^d)}$~{\small [fb]} 
  & $\delta^\perc_{\mu}$ \hspace*{-1em}
  & $\delta^\perc_{{\rm PDF}+\alpha_s+m_b}$ \hspace*{-1em}
  & $\delta^\perc_{\rm PDF}$ \hspace*{-1em}
  & $\delta^\perc_{\alpha_s}$ \hspace*{-1em}
  & $\delta^\perc_{m_b}$
    \\[0.7ex]
\hline
    \rule{0pt}{3ex}
  \hspace*{-1em} 4F \enskip \qquad $tH$ 
     & 45.90(7)
     & $^{+3.6}_{-6.3}$
     & $^{+2.3}_{-2.3}$
     & \scriptsize $ \pm 0.9 $ \normalsize
     & $^{+0.6}_{-0.9}$
     & $^{+2.0}_{-2.0}$
     & 46.67(8)
     & $^{+4.3}_{-6.1}$
     & $^{+3.2}_{-1.9}$
     & \scriptsize $ \pm 0.9 $ \normalsize
     & $^{+1.6}_{-0.4}$
     & $^{+2.6}_{-1.6}$
  \\[0.3ex] 
     \rule{0pt}{3ex} 
 $\bar tH$ 
     & 23.92(3)
     & $^{+4.2}_{-6.6}$
     & $^{+2.5}_{-2.7}$
     & \scriptsize $ \pm 1.4 $ \normalsize
     & $^{+1.6}_{-1.8}$
     & $^{+1.4}_{-1.5}$
     & 24.47(5)
     & $^{+4.4}_{-6.8}$
     & $^{+2.5}_{-2.3}$
     & \scriptsize $ \pm 1.4 $ \normalsize
     & $^{+1.4}_{-1.4}$
     & $^{+1.6}_{-1.2}$
  \\[0.3ex] 
     \rule{0pt}{3ex} 
 $tH + \bar tH$ 
     & 69.81(11)
     & $^{+3.2}_{-6.6}$
     & $^{+2.8}_{-2.5}$
     & \scriptsize $ \pm 0.9 $ \normalsize
     & $^{+1.6}_{-1.7}$
     & $^{+2.1}_{-1.6}$
     & 71.20(11)
     & $^{+4.3}_{-6.5}$
     & $^{+3.0}_{-2.4}$
     & \scriptsize $ \pm 0.9 $ \normalsize
     & $^{+2.0}_{-1.1}$
     & $^{+2.0}_{-1.9}$
  \\[0.7ex] 
\hline 
    \rule{0pt}{3ex} 
  \hspace*{-1em} 5F \enskip \qquad $tH$ 
     & 48.80(5)
     & $^{+7.1}_{-1.7}$
     & $^{+2.8}_{-2.3}$
     & \scriptsize $ \pm 1.0 $ \normalsize
     & $^{+1.7}_{-1.1}$
     & $^{+2.0}_{-1.8}$
     & 47.62(5)
     & $^{+7.4}_{-2.2}$
     & $^{+3.0}_{-2.4}$
     & \scriptsize $ \pm 1.0 $ \normalsize
     & $^{+1.6}_{-0.8}$
     & $^{+2.4}_{-2.0}$
  \\[0.3ex] 
     \rule{0pt}{3ex} 
 $\bar tH$ 
     & 25.68(3)
     & $^{+6.8}_{-2.0}$
     & $^{+3.4}_{-2.9}$
     & \scriptsize $ \pm 1.4 $ \normalsize
     & $^{+1.9}_{-1.5}$
     & $^{+2.5}_{-2.0}$
     & 25.07(3)
     & $^{+7.4}_{-2.1}$
     & $^{+3.2}_{-2.9}$
     & \scriptsize $ \pm 1.4 $ \normalsize
     & $^{+1.7}_{-1.8}$
     & $^{+2.4}_{-1.8}$
  \\[0.3ex] 
     \rule{0pt}{3ex} 
 $tH + \bar tH$ 
     & 74.80(9)
     & $^{+6.8}_{-2.4}$
     & $^{+3.0}_{-2.4}$
     & \scriptsize $ \pm 1.0 $ \normalsize
     & $^{+1.5}_{-1.1}$
     & $^{+2.4}_{-1.9}$
     & 72.79(7)
     & $^{+7.4}_{-2.4}$
     & $^{+2.9}_{-2.3}$
     & \scriptsize $ \pm 1.0 $ \normalsize
     & $^{+1.2}_{-1.4}$
     & $^{+2.4}_{-1.6}$
 \\[0.7ex] 
 \hline
\end{tabular}
 \caption{NLO cross sections and uncertainties for $pp\to tHq$, 
 $\bar tHq$ and ($tHq+\bar tHq$) at the 13-TeV LHC. 
 {\sc NNPDF2.3} PDFs have been used 
 ({\sc NNPDF2.1} for $m_b$ uncertainty in 5F). 
 The integration uncertainty in the last digit(s) (in parentheses) as
 well as the scale dependence and the combined ${\rm PDF}+\alpha_s+m_b$
 uncertainty in eq.~\eqref{eq:pdfasmb_totunc} (in $\perc$) are reported.
 The individual PDF, $\alpha_s$ and $m_b$ uncertainties are also
 presented as a reference.}
\label{tab:xsec_NNPDF23}
\end{table*}

\begin{table*}
\center
\begin{tabular}{r|llc|llc}
 \hline
    \rule{0pt}{3ex}
    $t$-channel
  & $\sigma_{\rm NLO}^{(\mu_0^s)}$~{\small [fb]} 
  & $\delta^\perc_{\mu+{\rm FS}}$ \hspace*{-1em}
  & $\delta^\perc_{{\rm PDF}+\alpha_s+m_b}$ 
  & $\sigma_{\rm NLO}^{(\mu_0^d)}$~{\small [fb]} 
  & $\delta^\perc_{\mu+{\rm FS}}$ \hspace*{-1em}
  & $\delta^\perc_{{\rm PDF}+\alpha_s+m_b}$ 
    \\[0.7ex]
\hline
    \rule{0pt}{3ex} 
  \hspace*{-1em} 4F+5F \enskip \qquad $tH$ 
     & 47.64(7)
     & \scriptsize $ \pm 9.7 $ \normalsize
     & $^{+2.9}_{-2.3}$
     & 47.47(6)
     & \scriptsize $ \pm 7.7 $ \normalsize
     & $^{+3.1}_{-1.8}$
  \\[0.3ex] 
     \rule{0pt}{3ex} 
 $\bar tH$ 
     & 24.88(4)
     & \scriptsize $ \pm 10.2 $ \normalsize
     & $^{+3.5}_{-2.6}$
     & 24.86(3)
     & \scriptsize $ \pm 8.3 $ \normalsize
     & $^{+3.3}_{-2.3}$
  \\[0.3ex] 
     \rule{0pt}{3ex} 
 $tH + \bar tH$ 
     & 72.55(10)
     & \scriptsize $ \pm 10.1 $ \normalsize
     & $^{+3.1}_{-2.4}$
     & 72.37(10)
     & \scriptsize $ \pm 8.0 $ \normalsize
     & $^{+2.9}_{-2.3}$
  \\[0.7ex] 
 \hline
\end{tabular}
 \caption{Same as table~\ref{tab:xsec_NNPDF23}, but for the
 flavour-scheme combined results, according to
 eq.~\eqref{eq:enevelope_MU-FS}.} 
\label{tab:comb_xsec_NNPDF23}
\end{table*}

\begin{figure}
\center 
 \includegraphics[width=1\columnwidth]{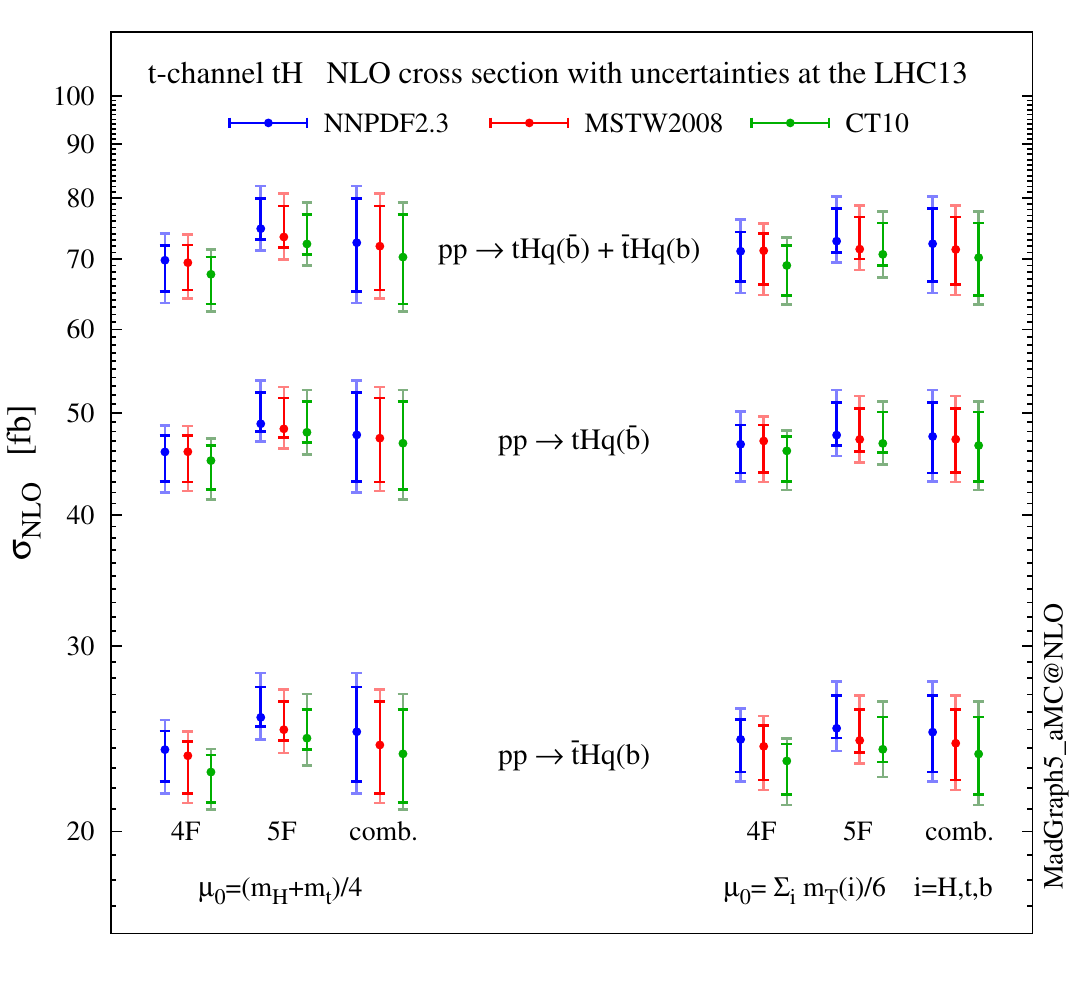}
 \caption{Summary plot of the NLO cross sections with uncertainties for
 Higgs production associated with a single top quark, via a $t$-channel $W$
 boson, at the 13-TeV LHC.
 For the uncertainties, the inner ticks display the scale (plus combined
 flavour-scheme) dependence $\delta_{\mu(+{\rm FS})}$, while the outer
 ones include $\delta_{{\rm PDF}+\alpha_s+m_b}$.} 
\label{fig:xsect_withunc}
\end{figure}

In table~\ref{tab:xsec_NNPDF23}, 
we report the NLO cross sections and their uncertainties at the 13-TeV LHC,
for $t$-channel $tH$ and $\bar tH$ productions separately, and for their sum $tH+\bar tH$.
Results are shown, using NNPDF2.3, in the 4F and 5F scheme for the static and dynamic
scale choices in eqs.~\eqref{eq:mu0fix} and \eqref{eq:mu0dyn},
including the sources of uncertainty discussed above: 
scale uncertainty and combined PDF, $\alpha_s(m_Z)$ and $m_b$ one as
well as the individual ones. 
The predictions in the combination of the 4F and 5F schemes defined in
eq.~\eqref{eq:enevelope_MU-FS} are presented in table~\ref{tab:comb_xsec_NNPDF23}. 
The theoretical uncertainty is dominated by the combined scale and
flavour-scheme uncertainty $\delta_{\mu+{\rm FS}}$ over the PDF,
$\alpha_s$ and $m_b$ uncertainty $\delta_{{\rm PDF}+\alpha_s+m_b}\,$.
Figure~\ref{fig:xsect_withunc} summarises the NLO cross sections and the
theoretical uncertainties for $t$-channel $tH$ production, including the
MSTW2008 and CT10 predictions. 

We conclude this section by commenting on two additional minor sources of uncertainty.
The first one is related to the value of the Higgs and top quark masses. 
In table~\ref{tab:mhmtvar_5F} we collect results for the $t$-channel NLO cross section 
(in the 5F scheme only) with parametric variations of 1~GeV in $m_H$ and $m_t$. 
The variations have a modest impact on the total cross section, about $1\perc$ only when both masses are varied in the same direction.
From the combination of Tevatron and LHC experimental results~\cite{ATLAS:2014wva} the top mass 
is currently known with a precision better than 1~GeV, while the
combination of the latest ATLAS and CMS
measurements of the Higgs 
mass gives a precision better than 0.5~GeV~\cite{Aad:2015zhl}.
We conclude that the impact of these uncertainties on the $t$-channel cross section 
at the LHC is below $1\perc$. 
The last source of uncertainty we discuss is the Yukawa coupling of the bottom quark.
We have checked that it is completely negligible, both in the 4F and 5F schemes, 
the impact of turning $y_b$ on/off at NLO being smaller than the numerical accuracy ($0.1-0.2 \perc$).
Finally, we remind the reader that EW corrections for this process are presently unknown, and these could have an impact on the accuracy of the present predictions.

\begin{table}
\centering
\begin{tabular}{r|lll}
 \hline 
   \rule{0pt}{3ex} 
     & & $m_t$ & \\
   $\sigma_{\rm NLO}^{({\rm 5F}\,\mu_0^s)}$~{\small [fb]} & 172.3    & 173.3    & 174.3
   \\[0.7ex]
 \hline 
   \rule{0pt}{3ex} 
   124.0
 & 75.54  $\scriptstyle (+1.0\perc)$ 
 & 75.18  $\scriptstyle (+0.5\perc)$
 & 74.99  $\scriptstyle (+0.3\perc)$ 
   \\[0.3ex]
   \rule{0pt}{3ex} 
   $m_H$ 125.0
 & 75.10  $\scriptstyle (+0.4\perc)$
 & 74.80               
 & 74.43  $\scriptstyle (-0.5\perc)$
   \\[0.3ex]
   \rule{0pt}{3ex} 
   126.0
 & 74.70  $\scriptstyle (-0.1\perc)$
 & 74.16  $\scriptstyle (-0.8\perc)$
 & 73.74  $\scriptstyle (-1.4\perc)$ 
   \\[0.7ex]
 \hline
\end{tabular}
 \caption{Higgs and top quark mass dependence of the NLO cross sections
 in the 5F scheme for $pp\to tHq+\bar tHq$ at the LHC with
 $\sqrt{s}=13$~TeV. 
 NNPDF2.3 PDFs have been used with $\mu_0=(m_H+m_t)/4$.  
 The figures in parentheses are the $\perc$ variations with respect to
 the reference cross section, computed with $m_H=125.0$~GeV and
 $m_t=173.3$~GeV.} 
\label{tab:mhmtvar_5F}
\end{table}

%%%%%%%%%%%%%% Begin Section: Distributions %%%%%%%%%%%%%%%%%%%%%%%%%%%%%%%%% 
\subsection{Distributions}\label{sec:distr}

\begin{figure*}
\center 
 \includegraphics[width=0.325\textwidth]{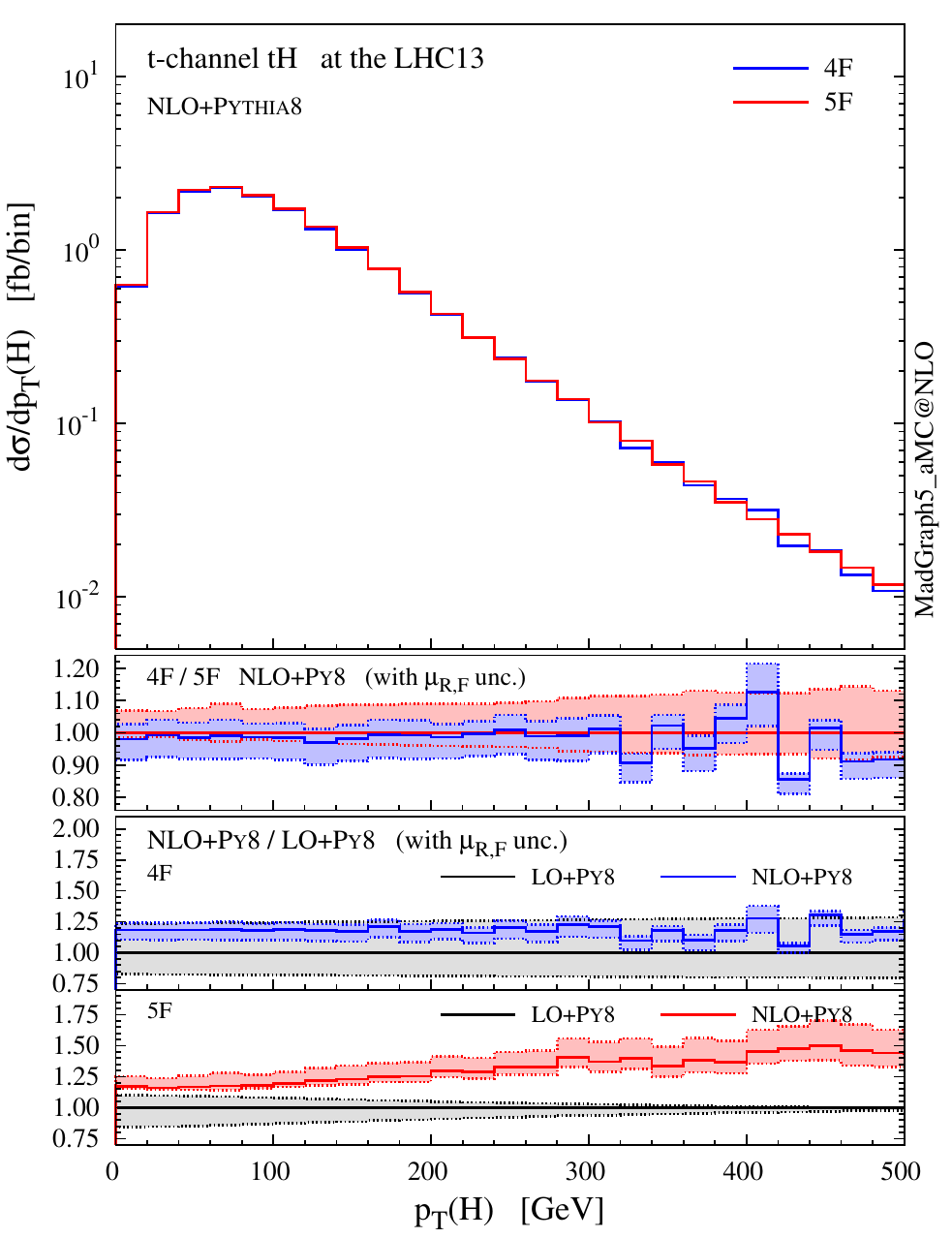}
 \includegraphics[width=0.325\textwidth]{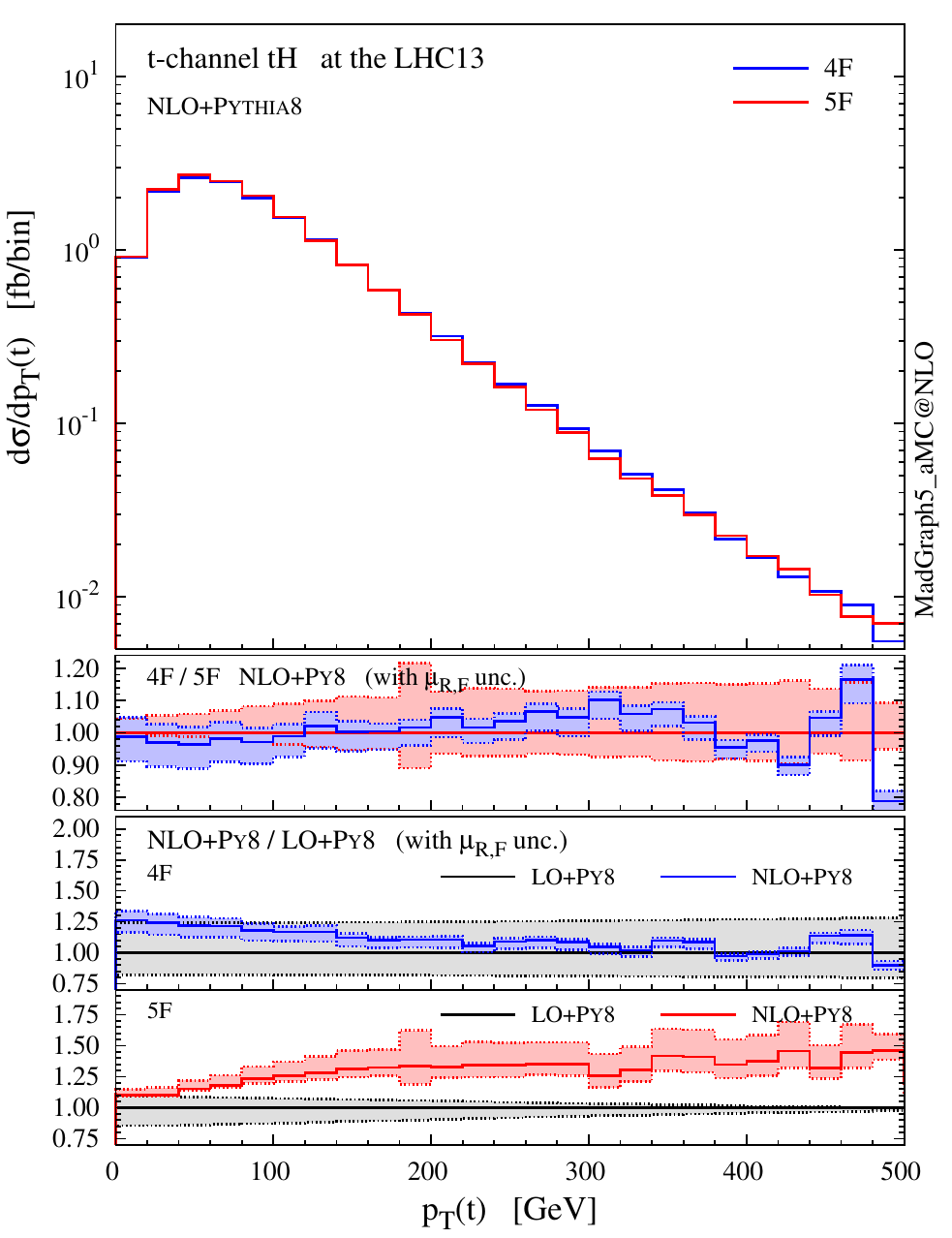}
 \includegraphics[width=0.325\textwidth]{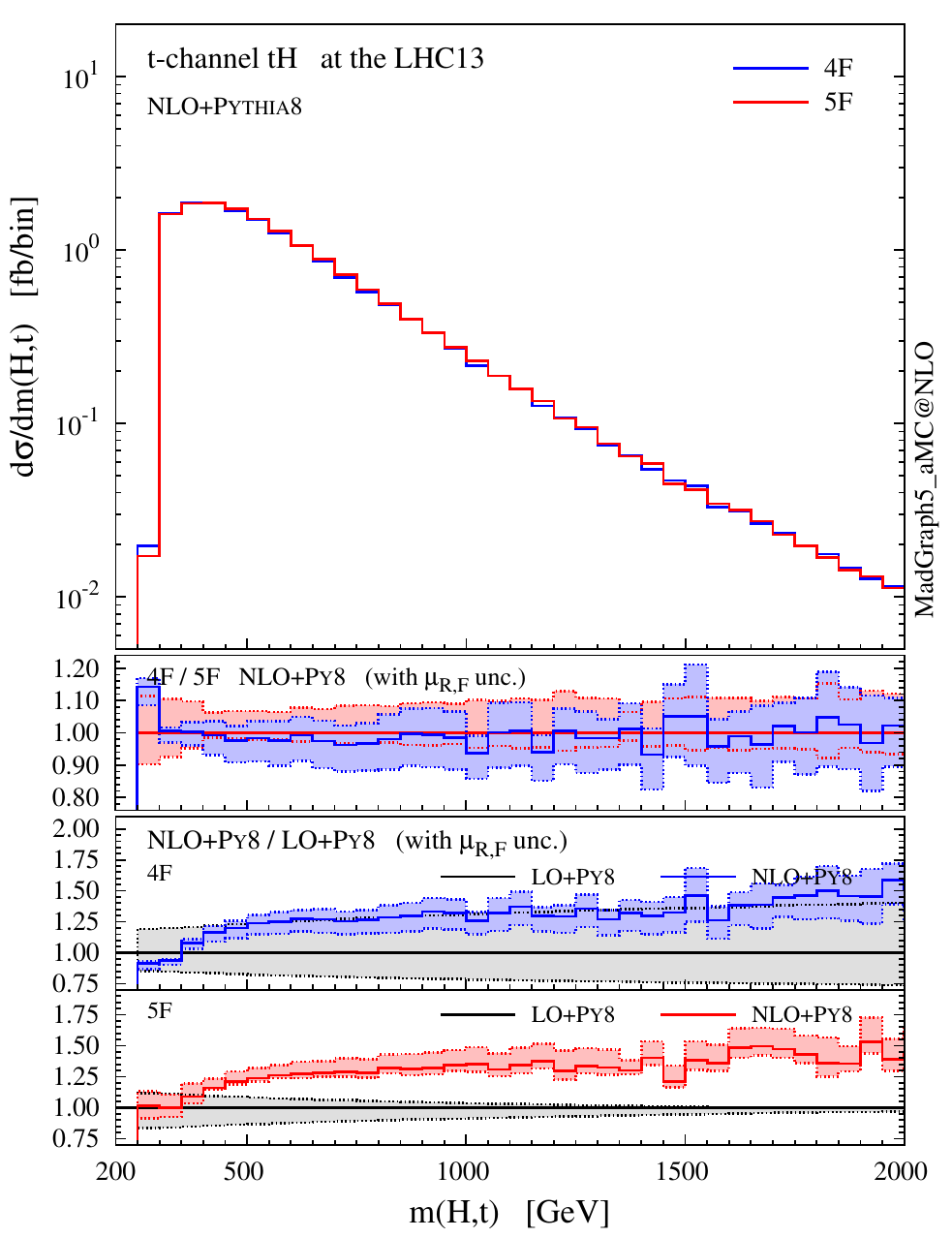}\\[1mm]
 \includegraphics[width=0.325\textwidth]{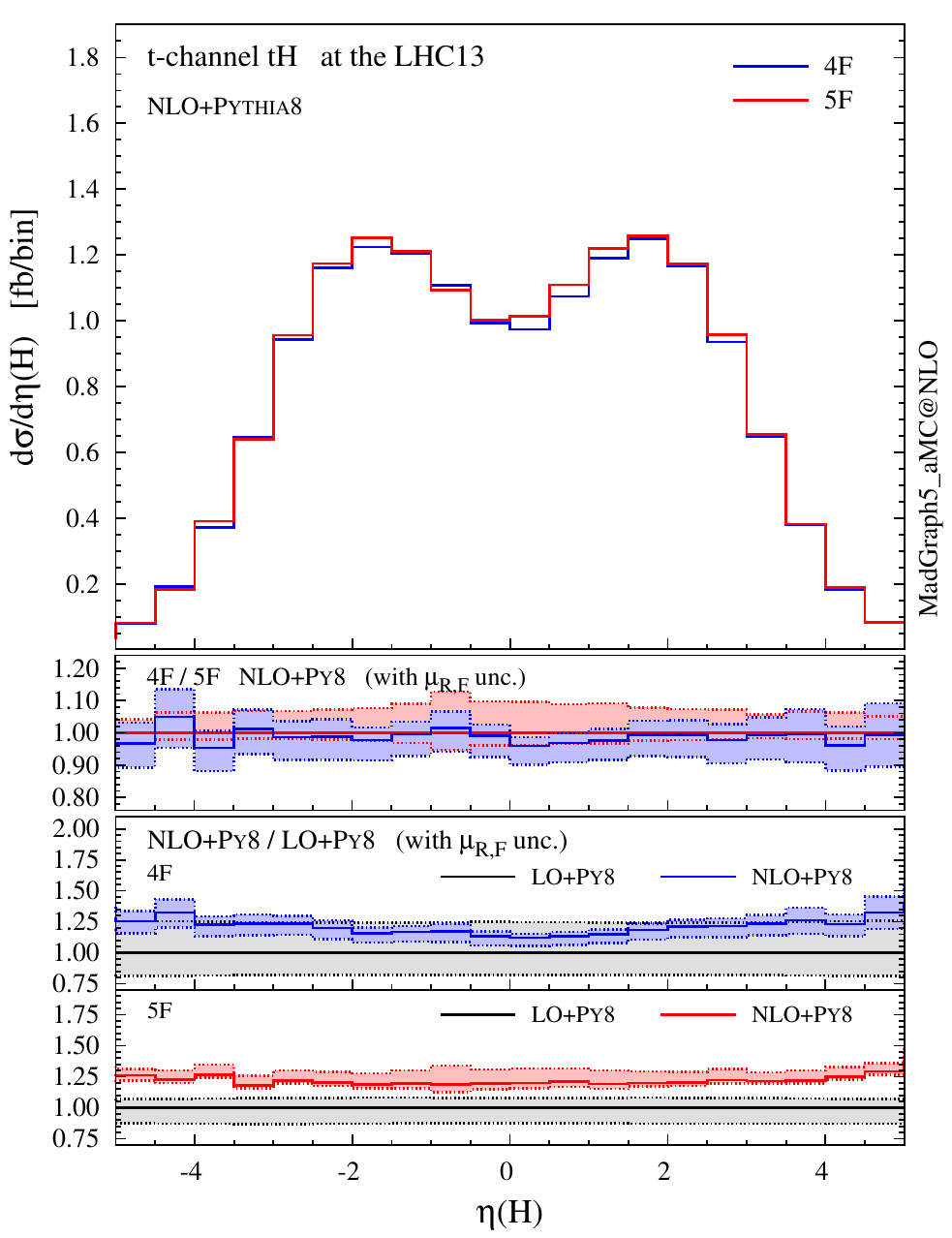}
 \includegraphics[width=0.325\textwidth]{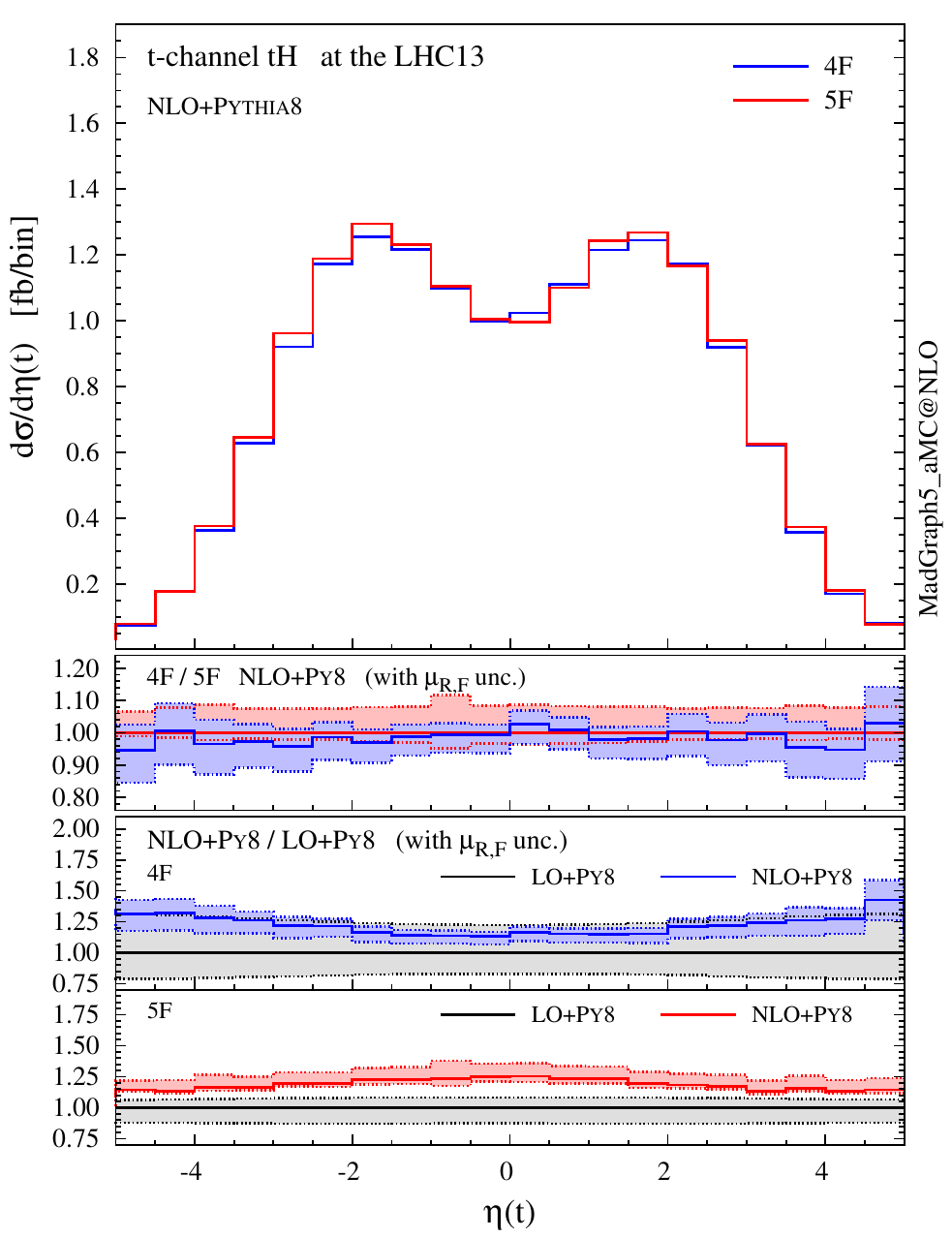}
 \includegraphics[width=0.325\textwidth]{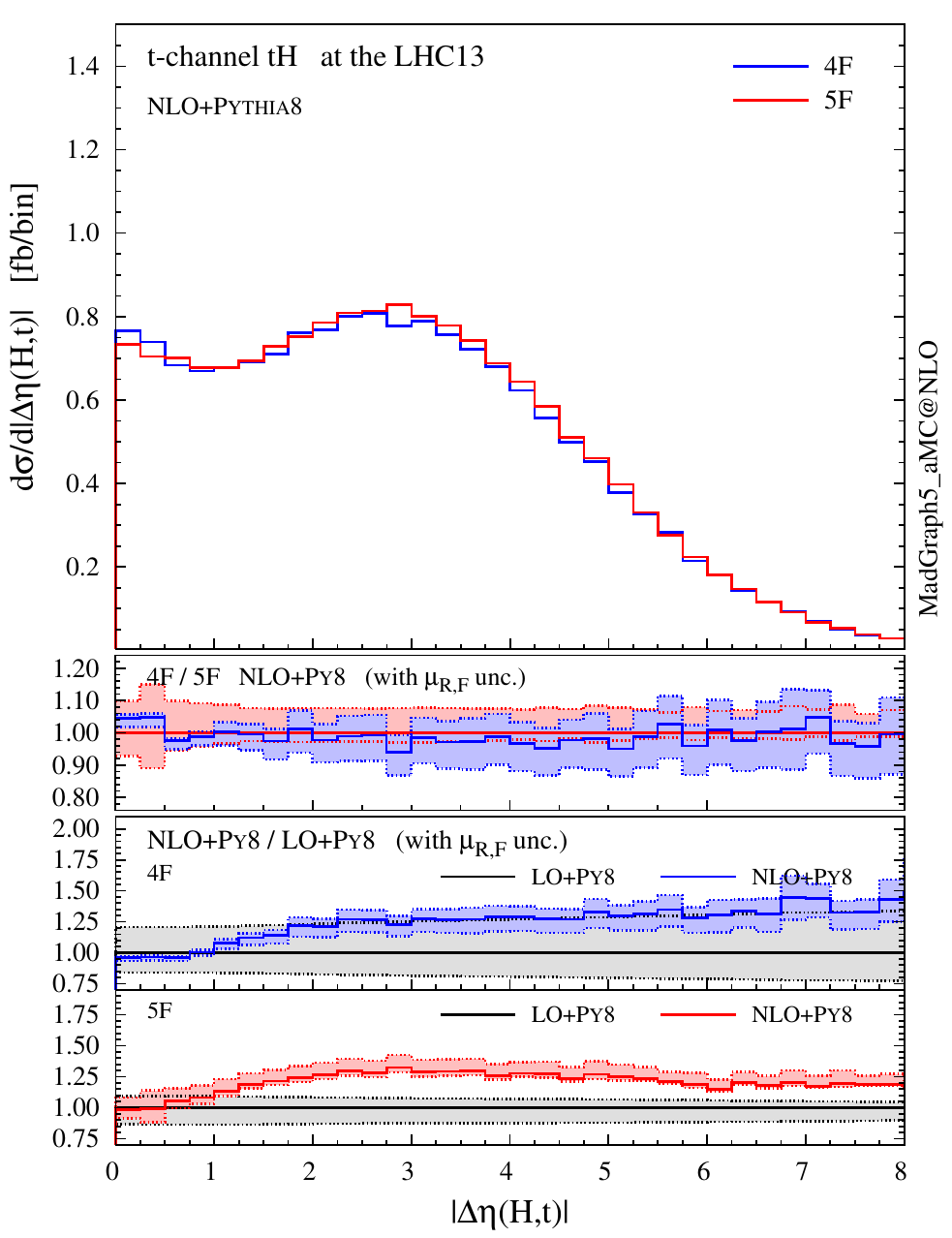}
 \caption{Representative differential distributions for the Higgs boson
 and the top quark at NLO+PS accuracy in $t$-channel $tH$ associated
 production at the 13-TeV LHC. 
 The lower panels provide information on the differences between 4F and
 5F schemes as well as the differential $K$ factors in the two schemes.}  
\label{fig:ht}
\end{figure*}

We now present a selection of kinematical distributions for the combined
$t$-channel $tH+\bar tH$ production at the 13-TeV LHC, with NLO
corrections and matching to a parton shower (NLO+PS). 
For the sake of brevity, we do not consider top and anti-top processes
separately in this section, and will dub with $t$ both the top quark and
its antiquark. 
Our main interest here is to assess the precision of the predictions for
$t$-channel production, therefore we do not specify any decay mode for
the Higgs boson, {\it i.e.} we leave it stable in the simulation.
On the other hand, we consider (leptonic) top decays, which allows us to
compare the distributions of $b$-jets coming from the hard scattering to
the ones coming from the top quark.   

For the kinematical distributions, we use {\sc NNPDF 2.3} PDFs and
the {\sc Pythia8} parton shower.
We have compared predictions obtained with the {\sc MSTW2008} and 
{\sc CT10} PDF sets and found no difference worth to report. 
We have also employed the {\sc HERWIG6} parton shower to verify that some
important conclusions on the difference of the radiation pattern between
4F and 5F schemes were not dependent on shower programs.  
We estimate the scale dependence by varying $\mu_R$ and $\mu_F$
independently by a factor two around the reference dynamic scale $H_T/6$
defined in eq.~\eqref{eq:mu0dyn}, which provides smaller scale
dependence than the static choice for differential distributions,
especially for the high-$p_T$ region.  

\begin{figure*}
\center 
 \includegraphics[width=0.325\textwidth]{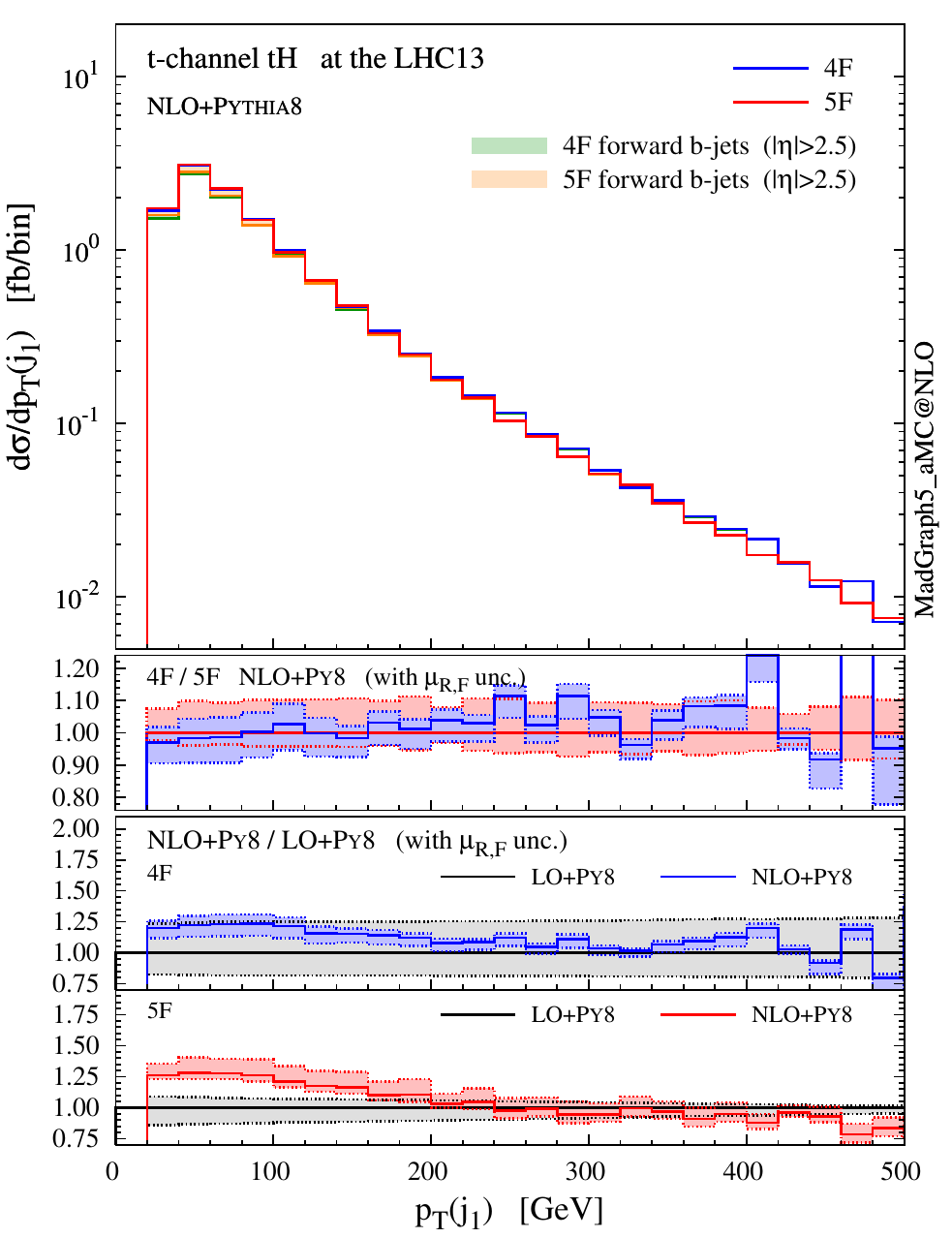}
 \includegraphics[width=0.325\textwidth]{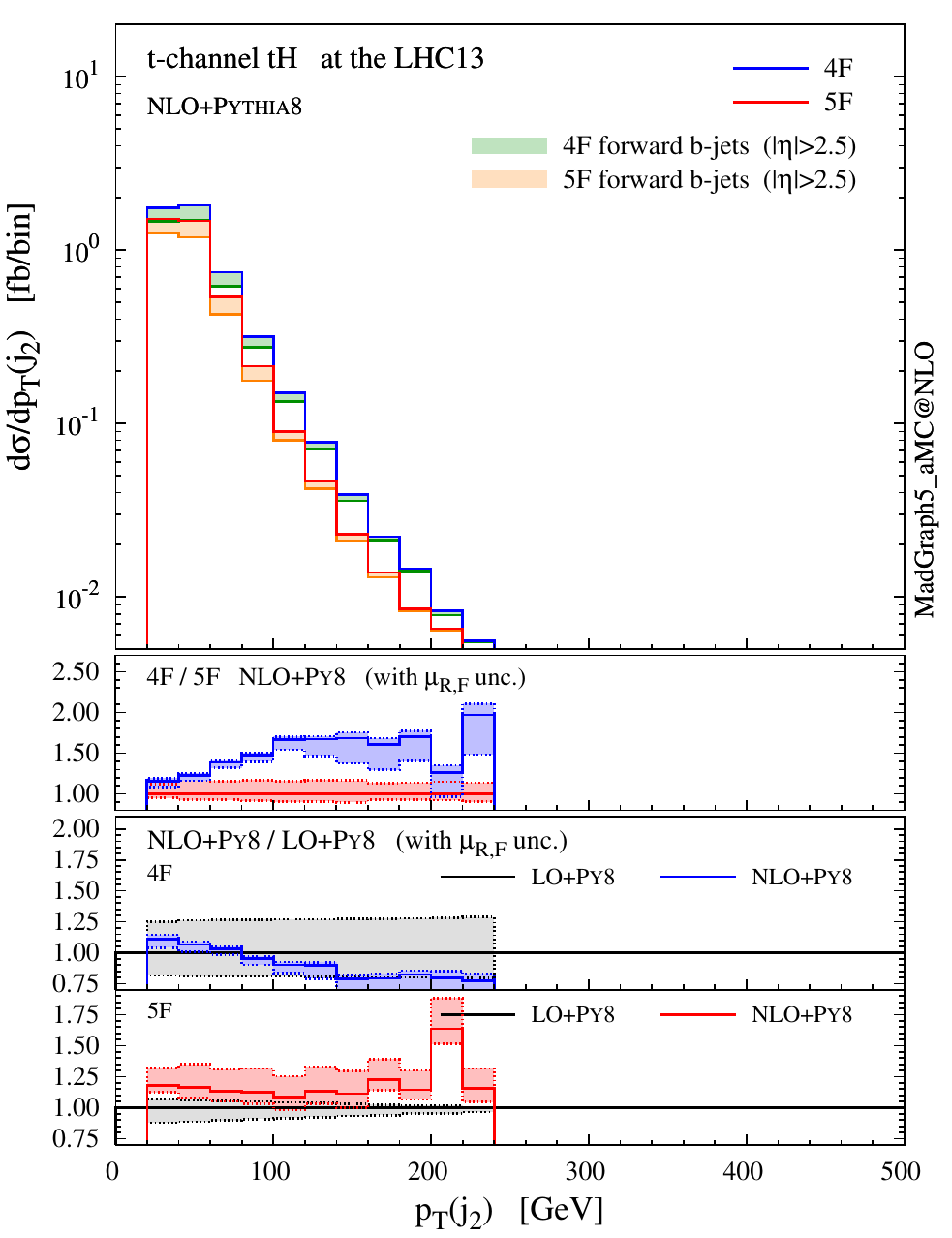}\\[1mm]
 \includegraphics[width=0.325\textwidth]{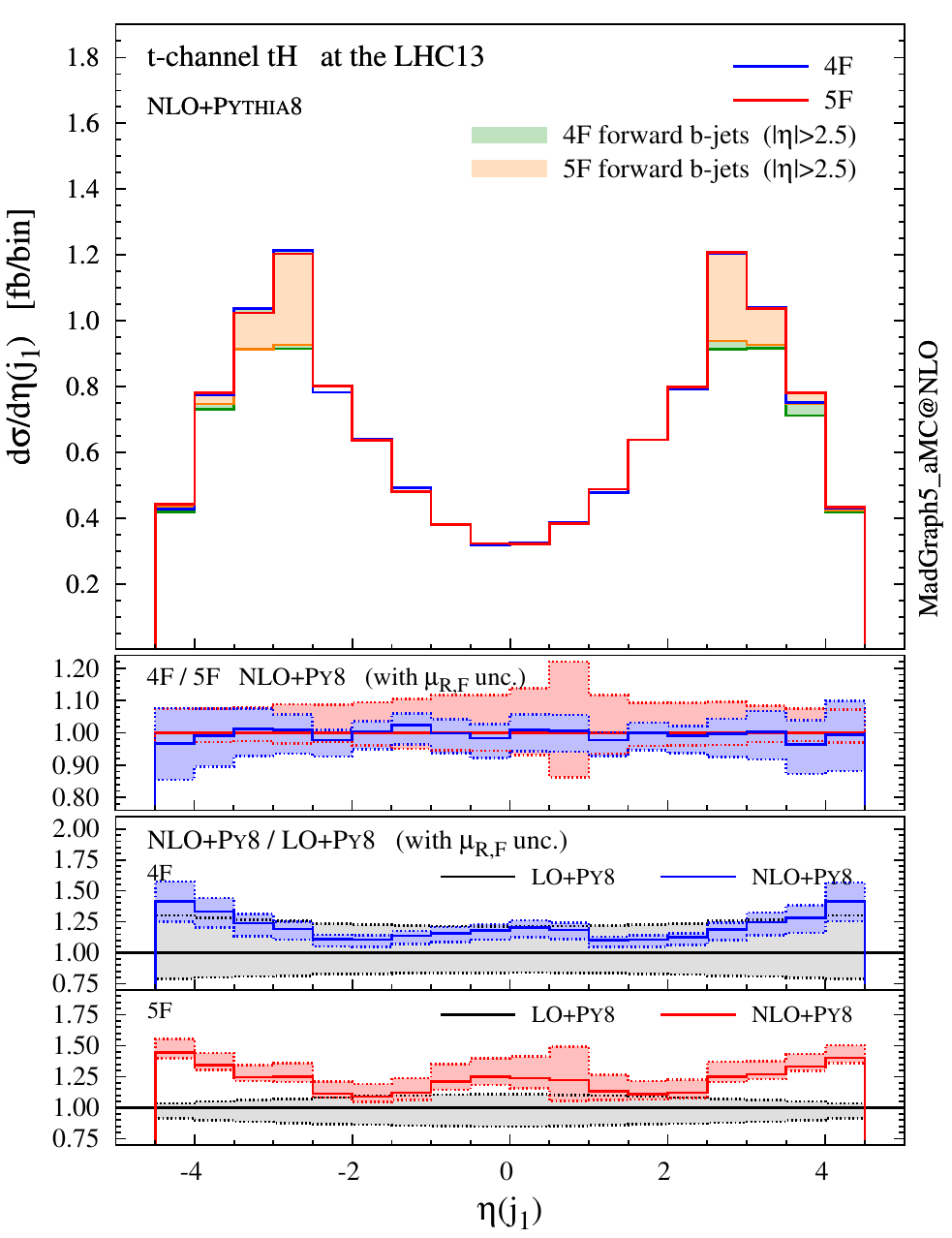}
 \includegraphics[width=0.325\textwidth]{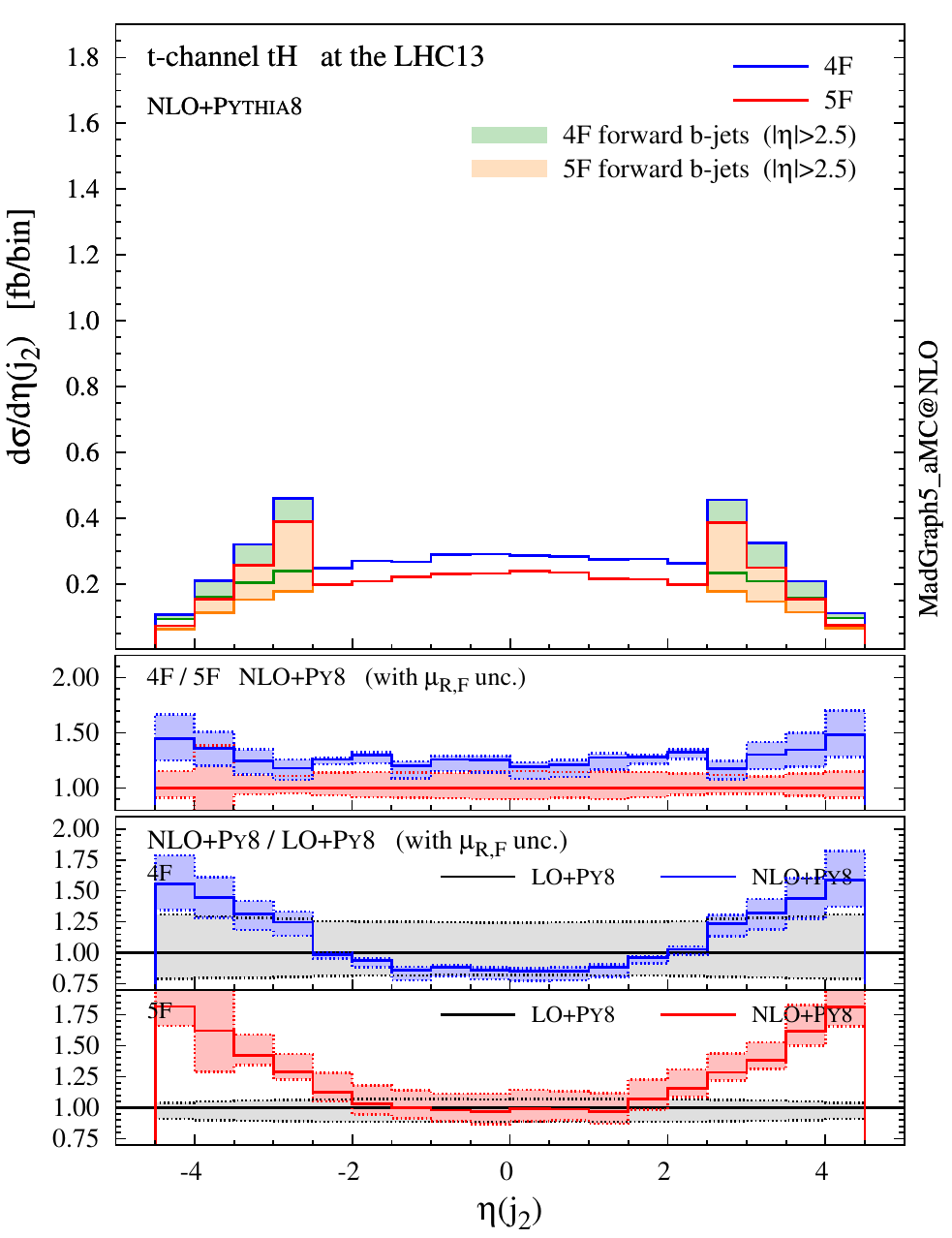}
 \caption{Same as fig.~\ref{fig:ht}, but for the two hardest jets.
 The contributions from non-taggable forward $b$-jets are also shown by
 shaded histograms as a reference.} 
\label{fig:jets}
\end{figure*} 

We start by showing in fig.~\ref{fig:ht} differential distributions for
the Higgs boson and the top quark (before they decay). 
The first observation is that NLO distributions in the 4F and 5F schemes
are in excellent agreement within their respective uncertainty associated to scale variation,
{\it i.e.} within the $10\perc$ level.  
Interestingly, though, differential $K$ factors (information in the
insets below) are more pronounced for the 5F than for the 4F scheme, the
NLO results in the 5F scheme typically being out of the uncertainties as
estimated from scale variation at LO.  
It should be noted that the LO process in the 5F scheme does not depend
on the renormalisation scale, and therefore its smaller uncertainty
(especially in the high-$p_T$ region) can be an artefact of the scheme. 
Results in the 5F tend to have a scale uncertainty that increases with $p_T$ much more than 
in the 4F, but in most cases the differences are not striking. 
Slightly larger deviations between 4F and 5F 
appear only very close to the $tH$ threshold, 
a region where we expect the 4F scheme to catch the underlying physics already at LO.

\begin{figure*}
\center 
 \includegraphics[width=0.325\textwidth]{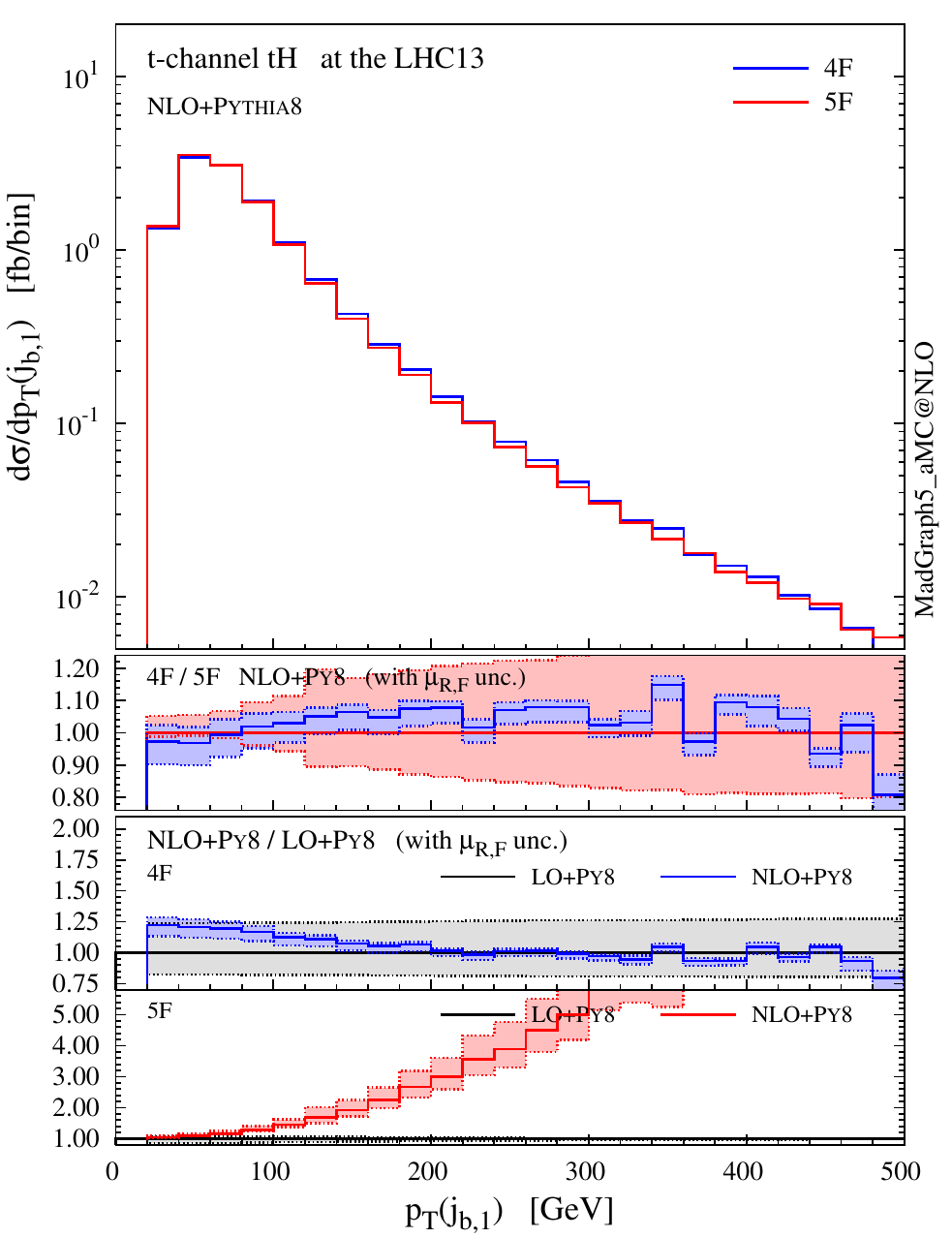}
 \includegraphics[width=0.325\textwidth]{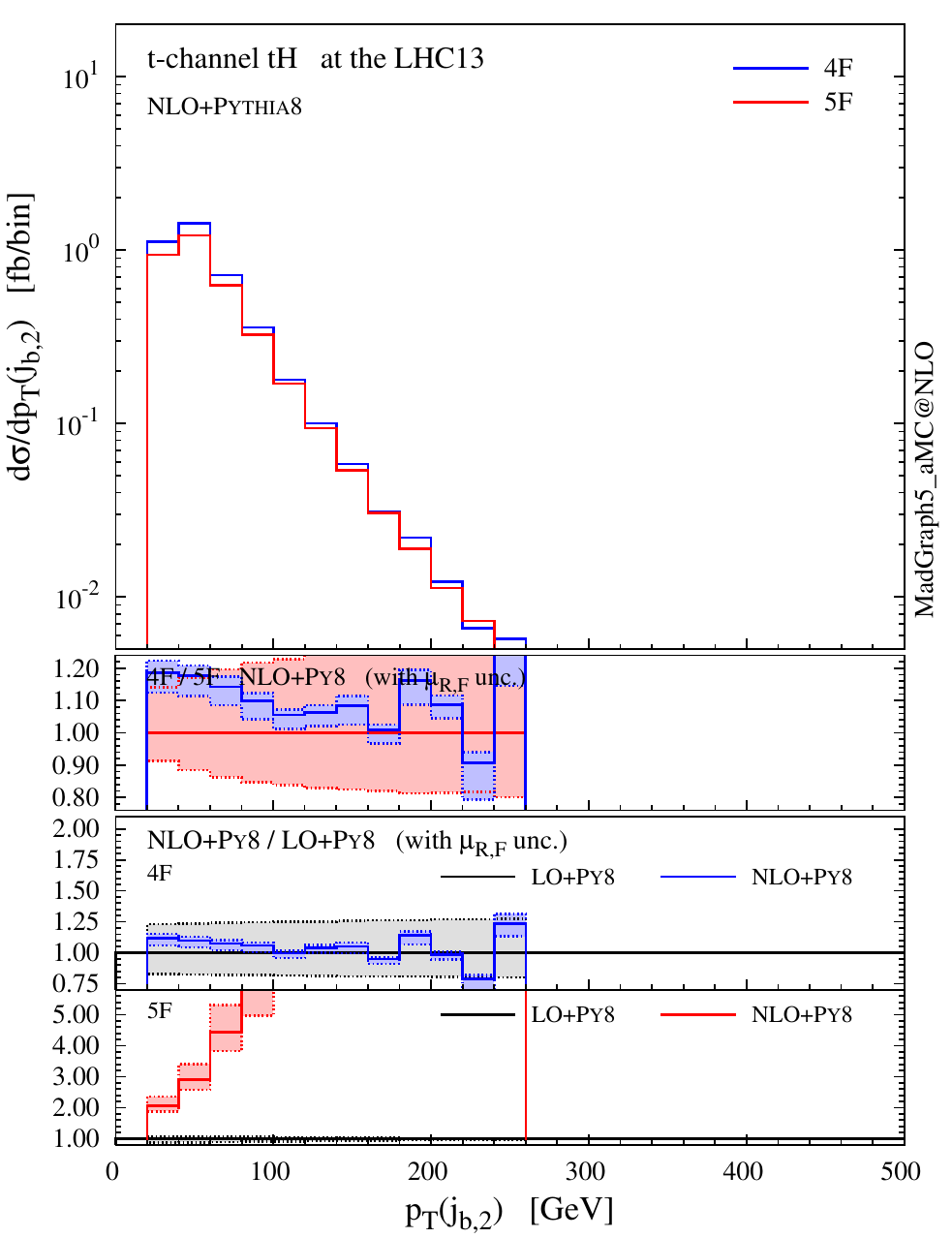}
 \includegraphics[width=0.325\textwidth]{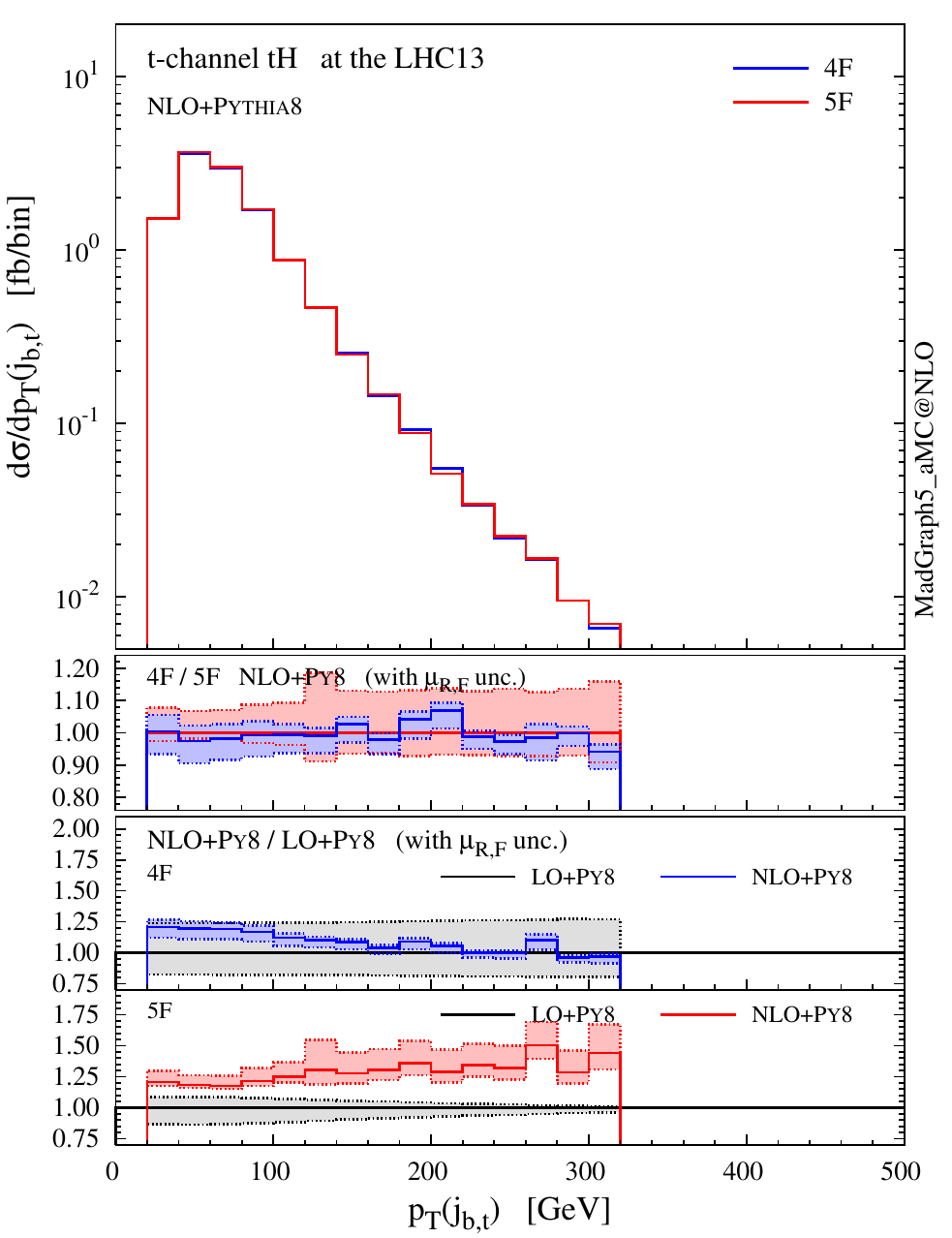}\\[1mm]
 \includegraphics[width=0.325\textwidth]{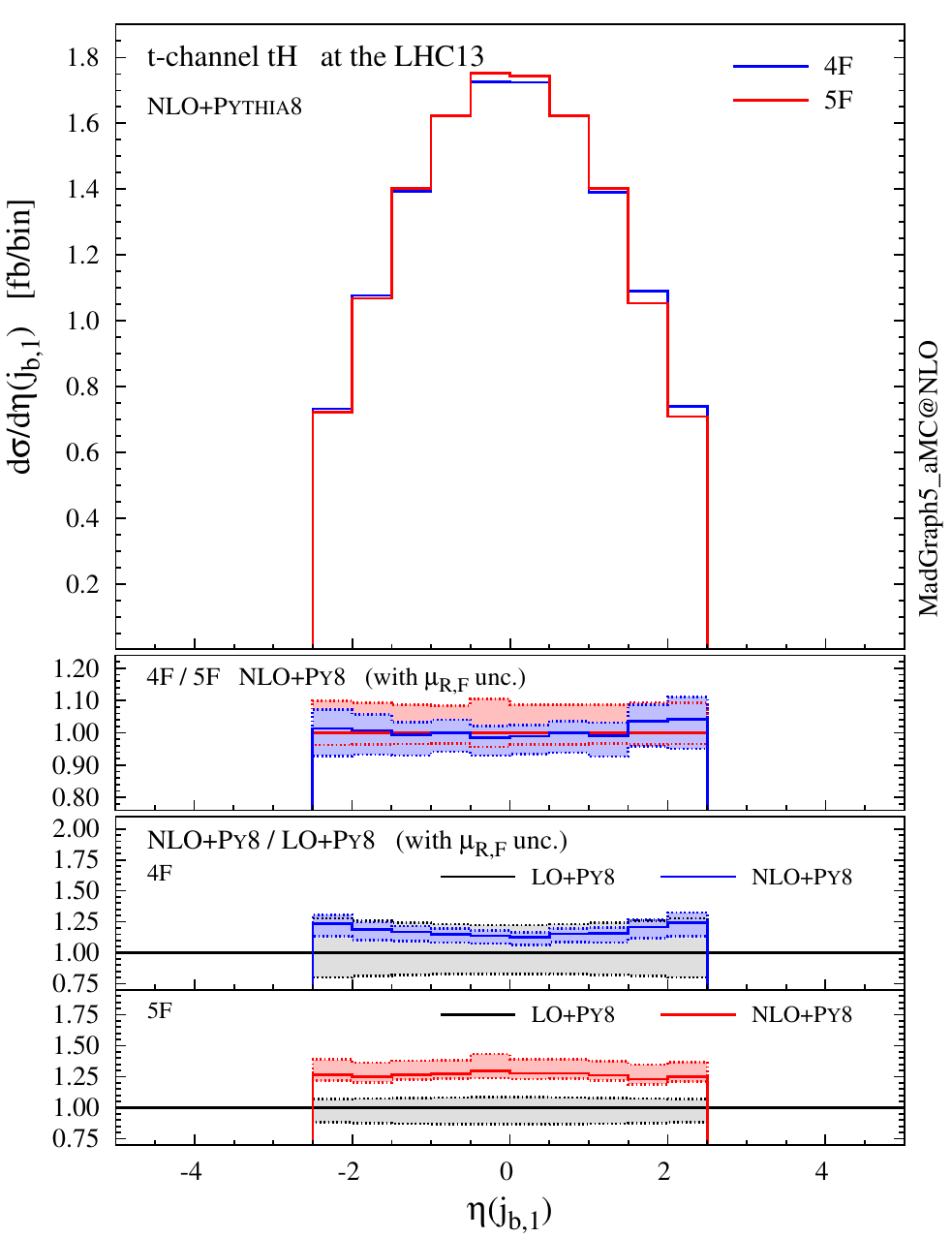}
 \includegraphics[width=0.325\textwidth]{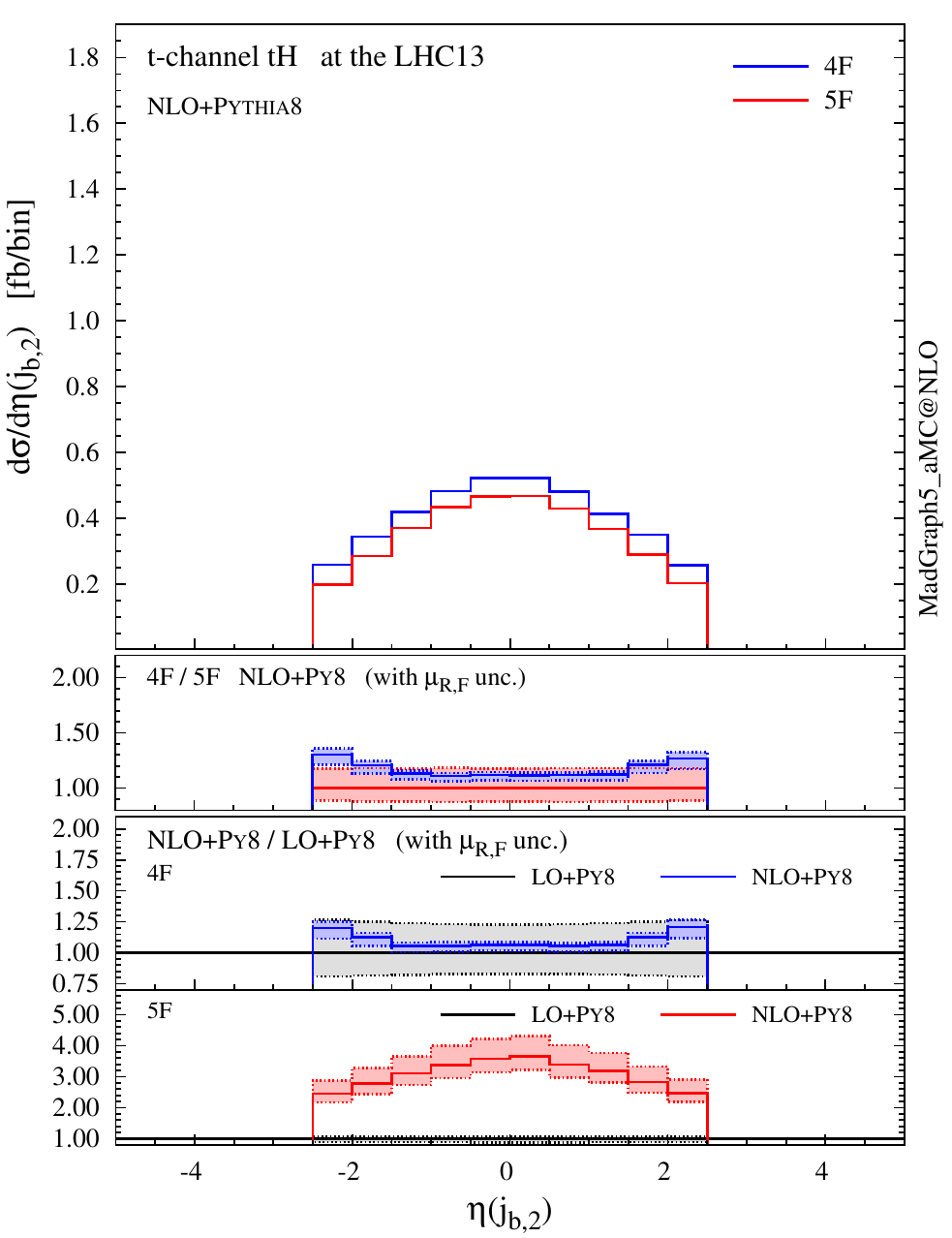}
 \includegraphics[width=0.325\textwidth]{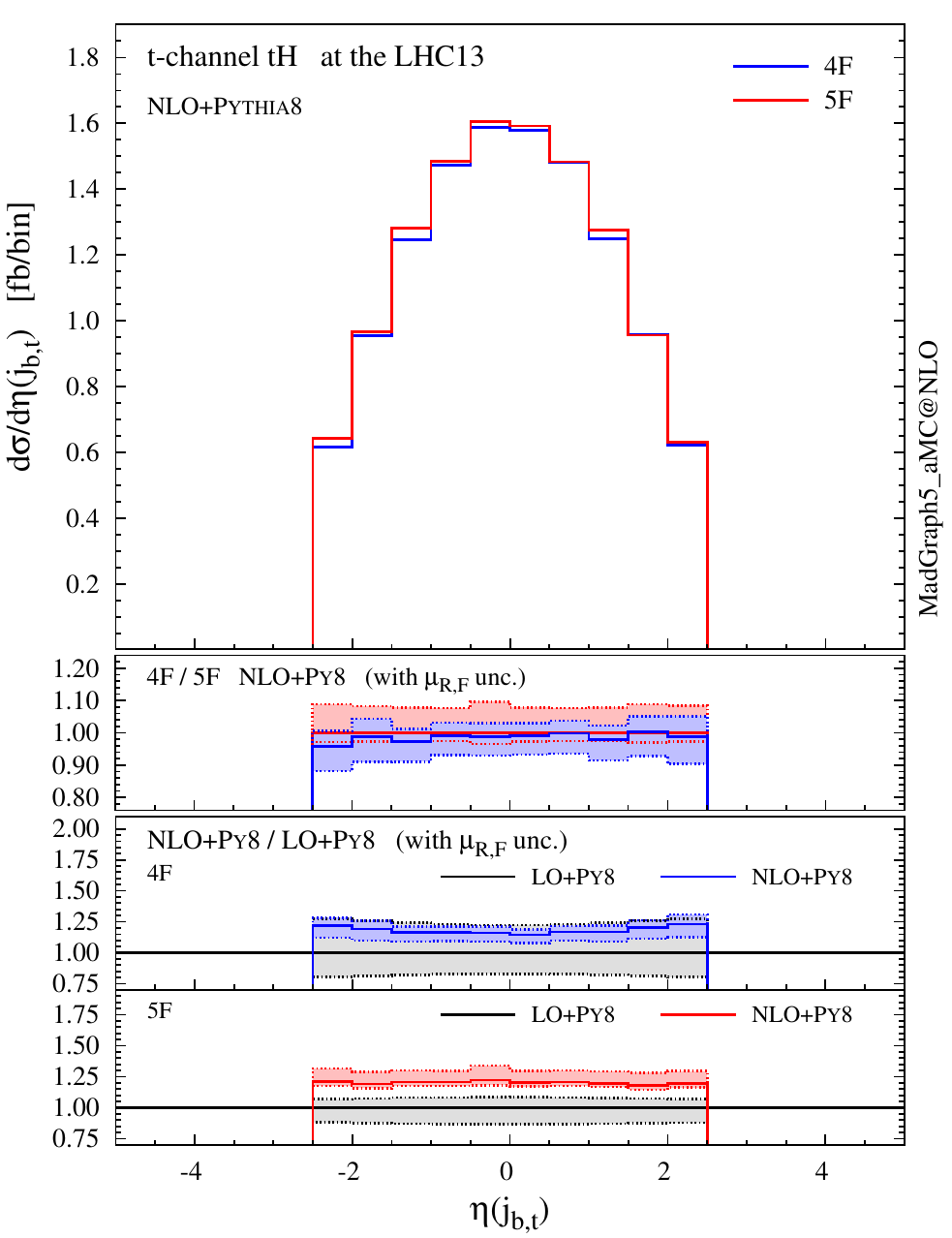}
\caption{Same as fig.~\ref{fig:ht}, but for the $b$-tagged jets.
 On the right column the distributions for the $b$-jet coming from the 
 top quark decay, selected by using Monte Carlo information, are shown.} 
\label{fig:bjets}
\end{figure*} 

In fig.~\ref{fig:jets} we present distributions for the two hardest jets
which are not tagged as $b$-jets.
Jets and $b$-jets are defined in eqs.~\eqref{eq:jet_definition} and
\eqref{eq:bjet_definition}.
The contributions from the non-taggable forward $b$-jets
($2.5<|\eta|<4.5$) are also denoted by shaded histograms as a reference.
The jet with the highest transverse momentum ($j_1$) tends to be
produced in the forward region, very much like in single-top and VBF
production. Most of the time this jet can be clearly associated to 
the light-quark current in the hard scattering. 
The very good agreement between 4F and 5F is manifest.
This is expected as this observable should not be too sensitive on the
details of heavy-quark current, as colour connections between the two
currents are either vanishing or suppressed at the order in QCD we are
working.   
On the other hand, sizeable differences arise for the second-hardest jet
($j_2$), which shows a much steeper $p_T$ spectrum and tends to be
produced centrally.
The difference between predictions in the 4F and 5F schemes is often
much larger than the scale uncertainty band (which is more pronounced in
the 5F scheme in the bulk of the events). 
We will discuss further this feature when presenting jet multiplicities
in the following. 

In fig.~\ref{fig:bjets} we show the analogous distributions for the
$b$-tagged jets.  
These are all the jets containing a $b$-hadron and falling inside the acceptance
of the tracking system, eq.~\eqref{eq:bjet_definition}.  
We consider the two hardest $b$-jets ($j_{b,1}$ and $j_{b,2}$) in the
event  regardless of their origin and, separately, we study the
$b$-jet coming from the top quark decay $j_{b,t}$ (tagged by using Monte
Carlo information). 
The $p_T$ spectrum of $j_{b,1}$ has a rather long tail compared to
$j_{b,2}$ and, at variance with light jets, all the $b$-jets tend to be
produced in the central region. 
Scale dependence at NLO is rather small in the 4F scheme, never reaching
$10 \perc$ and being typically around $5\perc$.  
Differences between 4F and 5F predictions are visible, specially in the
uncertainty band of $j_{b,2}$ in the 5F scheme;  
this is of course expected, given that this observable is described only
at LO accuracy in this scheme.  
Quite remarkably, however, these differences at NLO are often
significantly less pronounced than in the case of light jets (specially
for the second jet), while naively one might expect the $b$-jet
observables to be mostly affected by the flavour-scheme choice. 
On the other hand, at LO the inadequacy of the 5F scheme to describe
$b$-jets is evident. 

Comparing the transverse momentum of $j_{b,t}$ (first row, right plot in
fig.~\ref{fig:bjets}) to the corresponding spectra of $j_{b,1}$ and
$j_{b,2}$, it can be inferred that $b$-jets from the top quark mostly
contribute to the hardest $b$-jet ($j_{b,1}$) spectrum at low $p_T$.  
On the other hand, as the $p_T$ tail falls much more rapidly for
$j_{b,t}$ than for $j_{b,1}$, gluon splitting in the hard scattering
is the predominant mechanism at high $p_T$, and thus the main
source of $b$-jets in this region.  
This observation also explains why the scale dependence in the 5F is
small for low $p_T(j_{b,1})$, which is described at NLO accuracy, and
increases sharply in the high-$p_T(j_{b,1})$ region, 
where the physics is dominated by the transverse dynamics of the
$g\to b\bar b$ splitting, which is described only at LO.

\begin{figure}
\center 
 \includegraphics[width=0.325\textwidth]{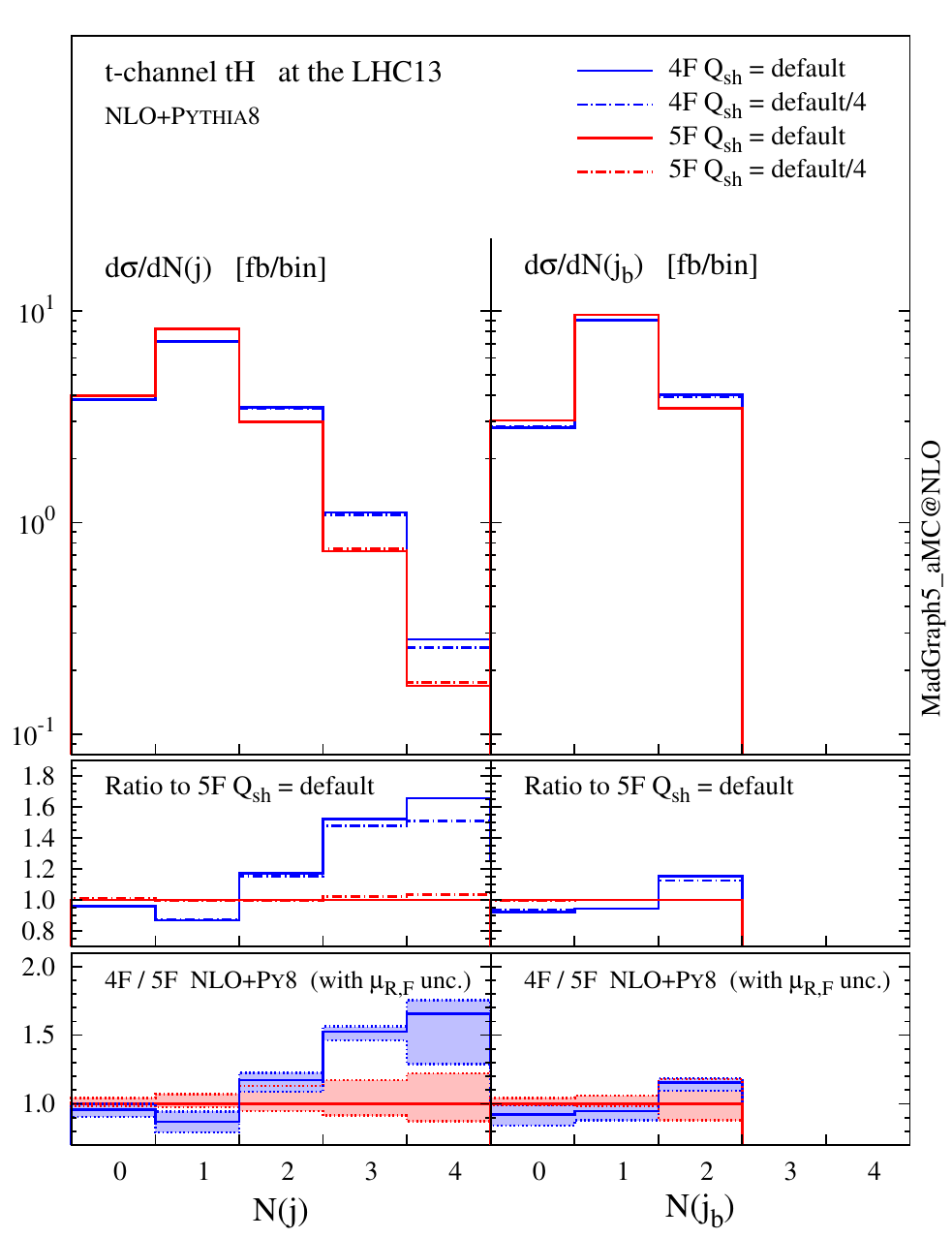}
 \caption{Jet rates at NLO+PS accuracy in 4F and 5F schemes with
 different choices of the shower scales.}
\label{fig:jetrate}
\end{figure} 

We conclude this section by studying the jet multiplicities, which are
sensitive to the flavour scheme as well as to the choice of the shower
scale. As argued in~\cite{Maltoni:2012pa}, the dynamics of $g\to b\bar b$
splitting takes place at a scale which is typically lower than the hard
scale of the process $m_t+m_H$ or $H_T$, affecting the choice for the
factorisation scale that one should use to describe $t$-channel
production.  
An analogous argument could be made also for the shower scale
choice~\cite{Wiesemann:2014ioa}, which in the 
{\sc MadGraph5\_aMC@NLO} matching procedure is chosen to
be of the order of the partonic centre-of-mass energy in the Born process.  
In fig.~\ref{fig:jetrate}, we study the dependence of jet rates on the
flavour scheme as well as on the shower scale, where two different choices
of the shower scale are compared:
one is the default value, and another is the default value divided by a
factor of four.  
We can see that reducing the parton-shower scale has only a minor impact
on the distributions, while a more interesting pattern arises from the
choice of the flavour scheme.  

For the $b$-tagged jets (right panel in fig.~\ref{fig:jetrate}),
differences between the two schemes are rather mild  ($\sim 15\perc$ in
the 2-jet bin and less for 0 and 1 jet) and always compatible within the
scale uncertainty, which for the 2-jet bin is much larger in the 5F (the
accuracy being only at LO). 

\begin{figure}
\center 
 \includegraphics[width=1\columnwidth]{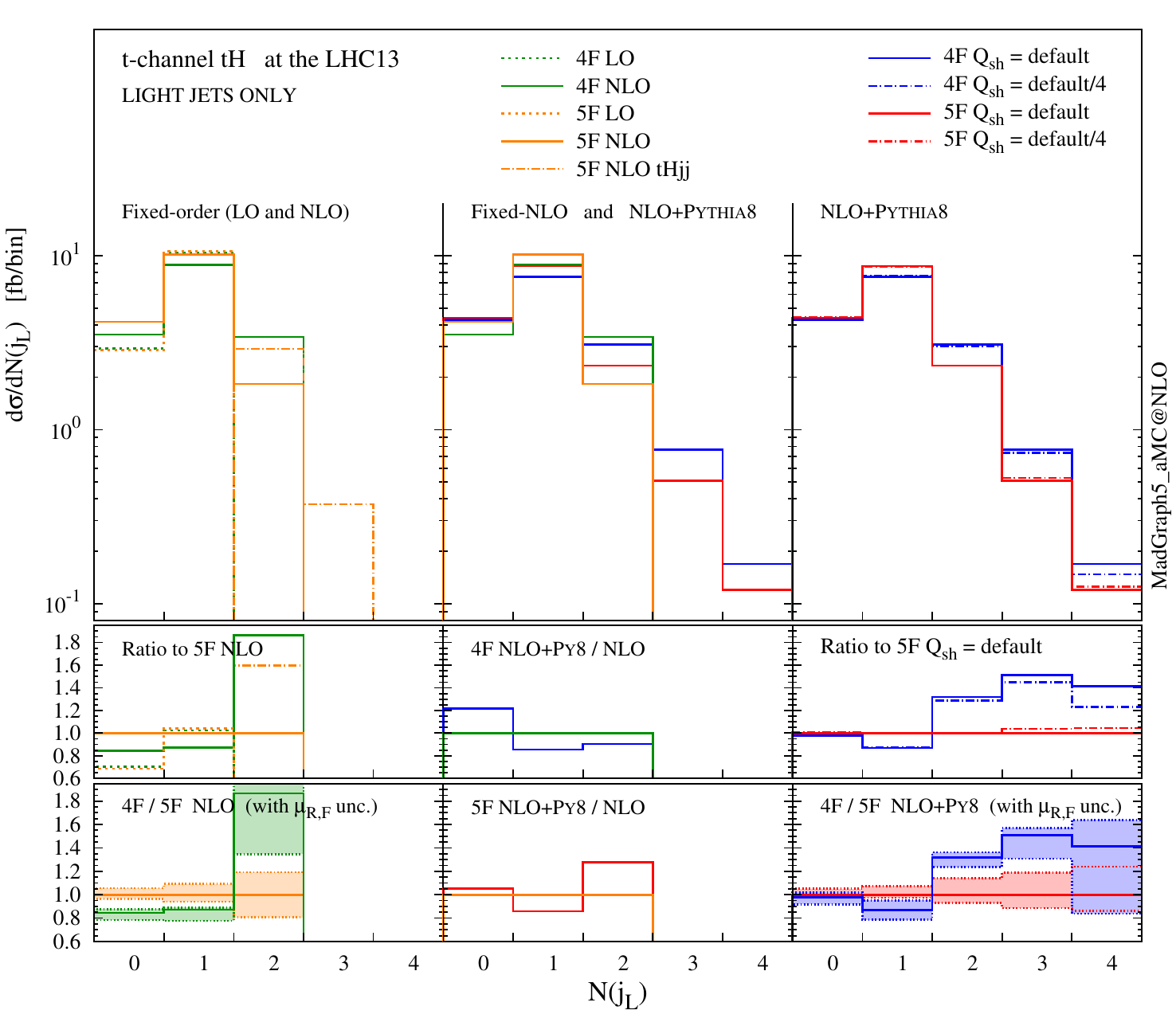}
 \caption{Jet rates only for the light jets both at fixed order and
 matched to a parton shower in 4F and 5F schemes with different choices
 of the shower scales.} 
\label{fig:jetmultiplicity}
\end{figure}

For non-$b$-tagged jets (left panel in fig.~\ref{fig:jetrate}), on the
other hand, a higher jet multiplicity is clearly observed in the 4F
scheme, which implies that harder QCD radiation is favoured in this
scheme. 
Interestingly, the difference is visible already at the 1-jet bin, which
is described at NLO accuracy at the matrix-element level.
These differences cannot arise from the small component of forward, non-taggable heavy jets; on the contrary, they can be 
understood by considering jets that come from genuinely light QCD radiation.
In fig.~\ref{fig:jetmultiplicity} we show explicitly
the multiplicity of light jets only (tagged by using
Monte Carlo information), both at fixed order in QCD and at NLO matched to parton shower.
Our first observation is that results in the 4F and 5F are almost identical at fixed LO 
(where only the zero and one jet bins are filled). The difference is therefore an effect of higher-order corrections, 
as it is confirmed by observing the fixed-NLO histograms.
We recall that the fixed-order matrix element has a different colour structure in different schemes;
in particular, the 4F at LO features a gluon in the initial state (compared to the $b$-quark in the 5F) 
and an extra $b$ in the final state. The radiation of  extra light QCD partons from the $g \to b \bar b$ splitting is therefore favoured in the 4F (e.g. an extra gluon can either attach to the initial-state gluon
or to one of the $b$'s, while in the 5F it can attach only to the initial-state $b$).
This is indeed what we observe at fixed NLO.

If the origin of the difference in the jet rates can  be traced back to the difference between the 
LO 4F and 5F colour structures, then one would also expect this
difference to be mitigated once higher-order corrections are included. To this aim, we have performed a fixed-order computation of the 2-jet bin in the 5F at NLO accuracy, {\it i.e.} calculated $tHjj$ at NLO, within our simulation framework, finding indeed that the  
rate is significantly enhanced (by $\sim 60\perc$), lying much closer to the 4F result.  A further hint that the scheme difference is indeed mitigated at higher orders is given by the NLO+PS results, which
show that the 2-jet bin in the 4F is reduced by $\sim 10 \perc$ after
the shower, while the corresponding 5F one is enhanced by $\sim 30
\perc$ over the fixed-order result. Finally, we have checked that the
same results we have found here for single top plus Higgs,  occur also
in the case of single top production alone. In conclusion, our results
suggest that the inclusion of the $g \to b \bar b$ splitting in the
matrix-element description at the lowest order, {\it i.e.} the 4F
scheme, allows a wider range of observables relevant for the analyses to
be described more accurately. 

%%%%%%%%%%%%%% Begin Section: s-channel %%%%%%%%%%%%%%%%%%%%%%%%%%%%%%%%%%%%
\section{$\boldsymbol{s}$-channel production}\label{sec:schannel}

Higgs-top quark associated production at hadron colliders can also be
mediated by $s$-channel diagrams, see fig.~\ref{fig:diagram-s}.
Compared to $t$-channel production, the $s$-channel mechanism is
naturally suppressed by the higher virtuality of the intermediate
$W$ boson and features a much smaller cross section at the LHC.
In this section we calculate the NLO cross section, evaluating the
corresponding uncertainties, and compare $s$-channel distributions to
those of $t$-channel production at NLO+PS level. 

\begin{figure*}
\center 
 \includegraphics[width=0.325\textwidth]{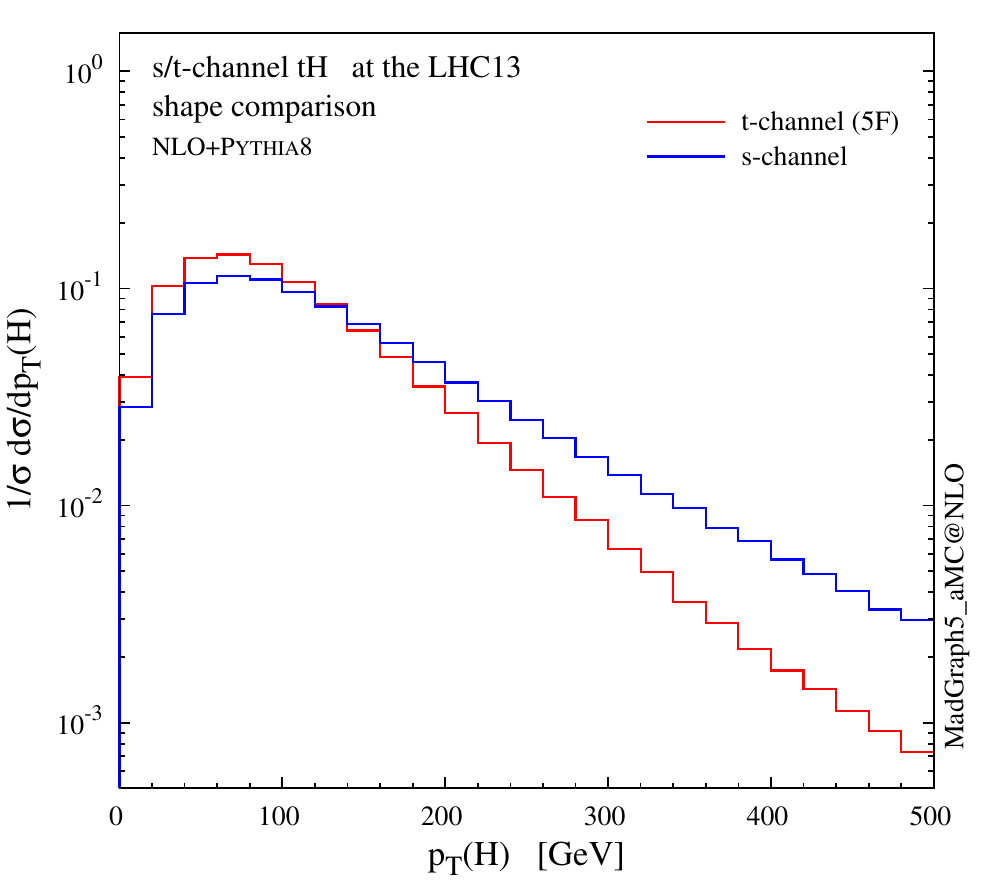}
 \includegraphics[width=0.325\textwidth]{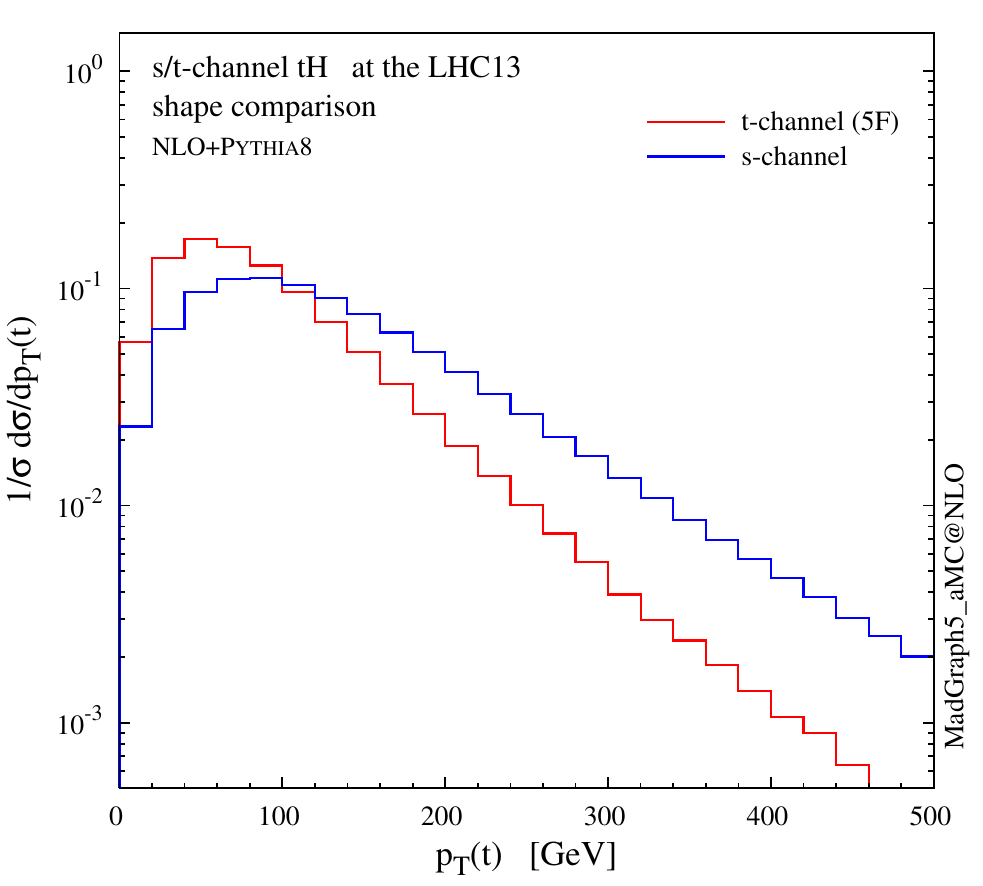}
 \includegraphics[width=0.325\textwidth]{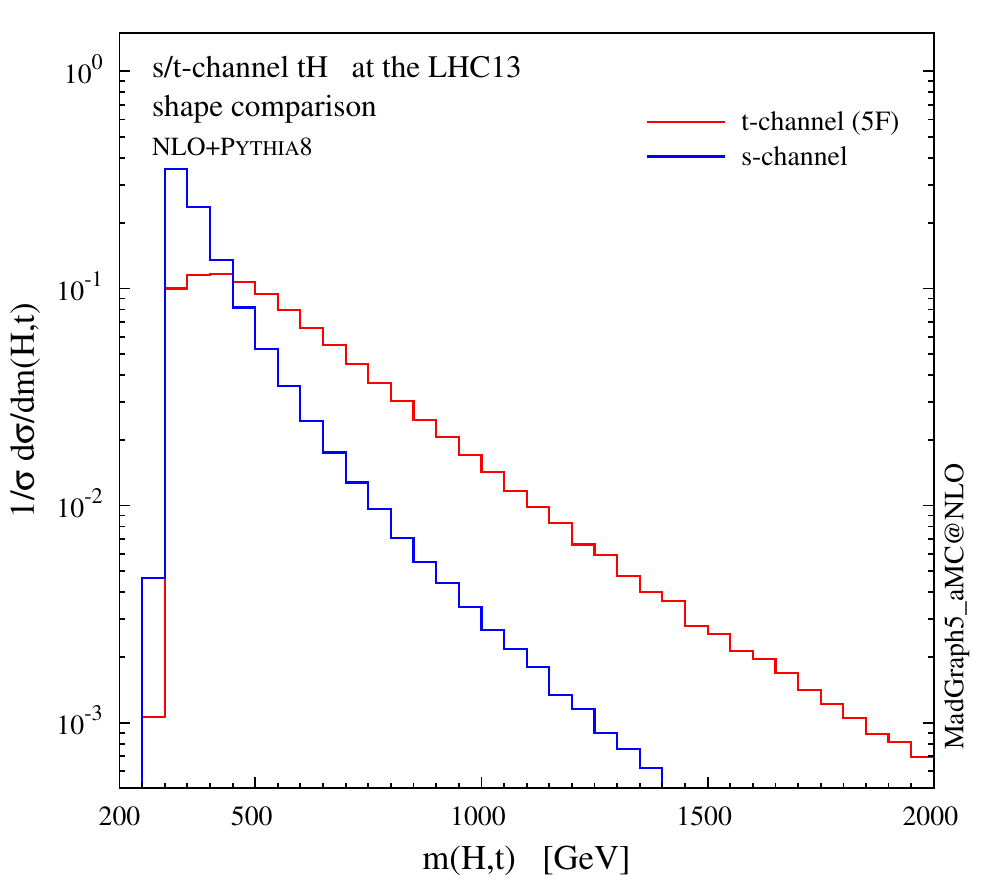}\\[1mm]
 \includegraphics[width=0.325\textwidth]{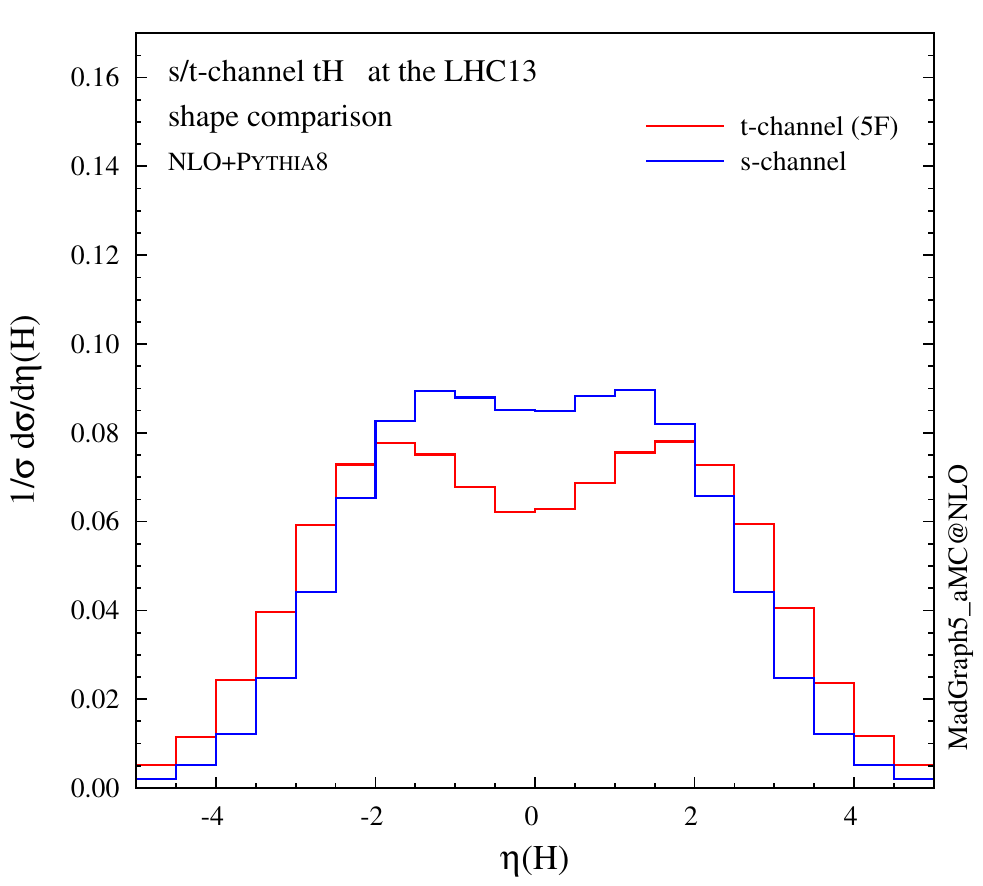}
 \includegraphics[width=0.325\textwidth]{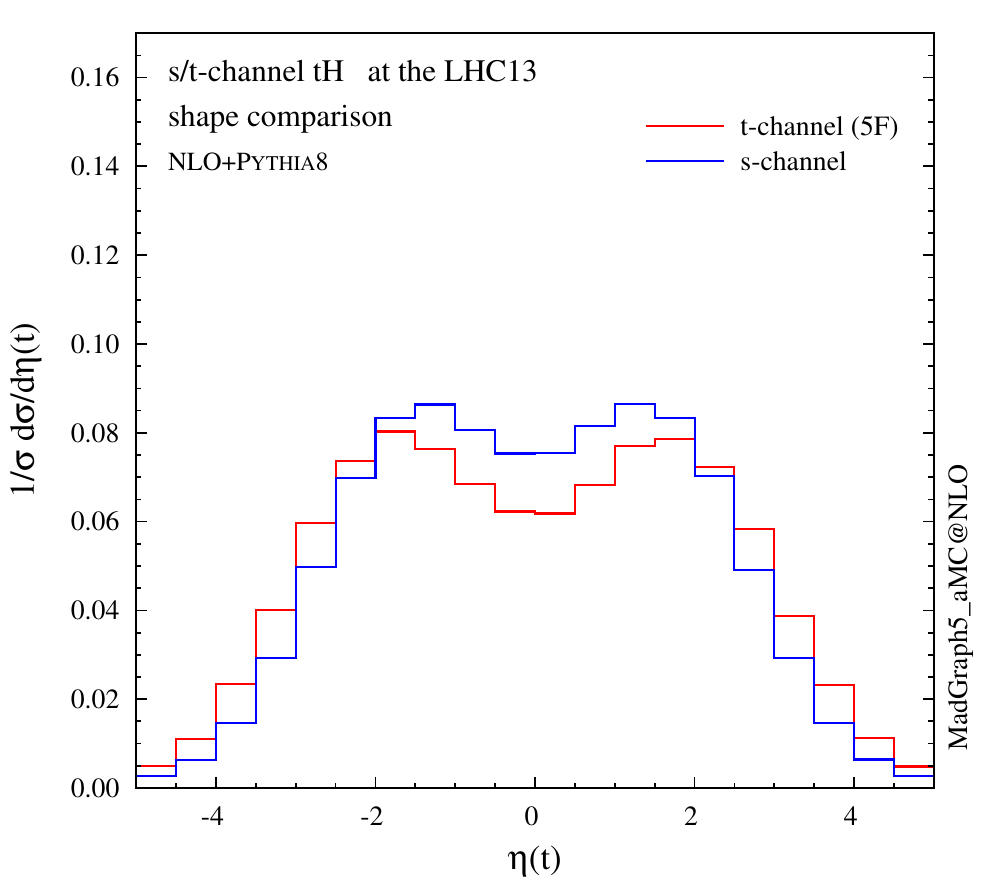}
 \includegraphics[width=0.325\textwidth]{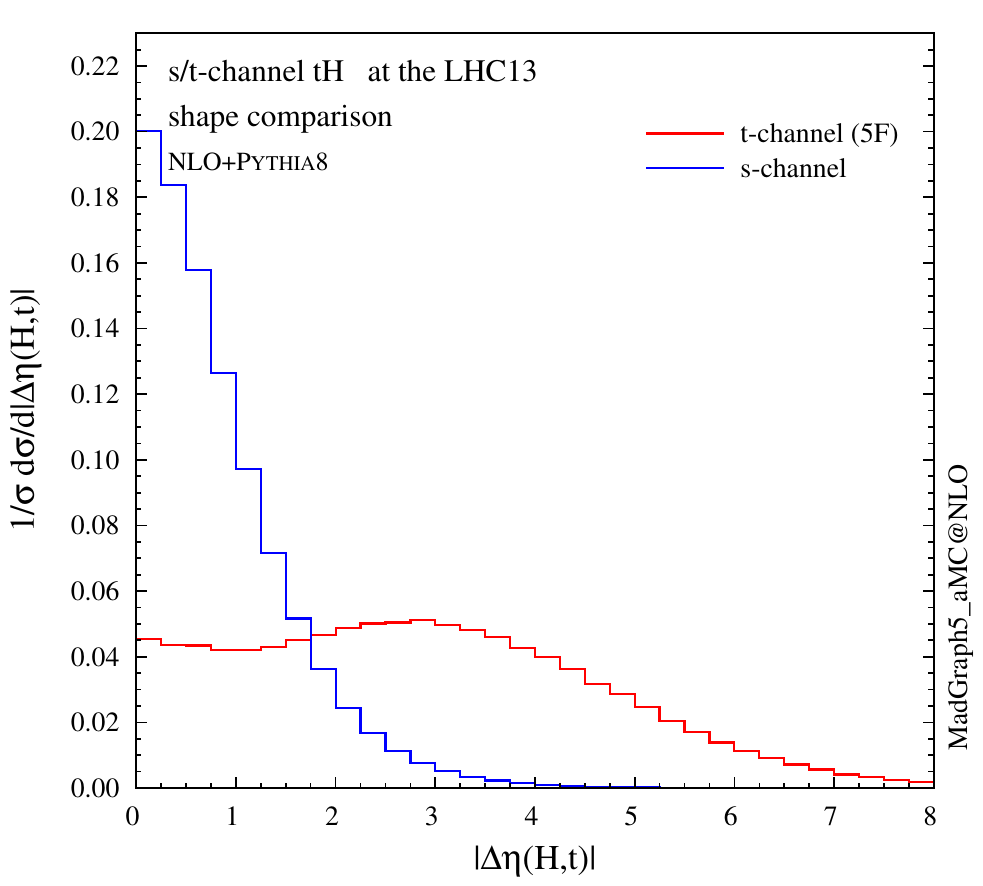}
 \caption{Shape comparison between $s$- and $t$-channel distributions
 for the Higgs boson and the top quark at NLO+PS accuracy.} 
\label{fig:STchannel_ht}
\end{figure*} 

\begin{figure*}
\center 
 \includegraphics[width=0.325\textwidth]{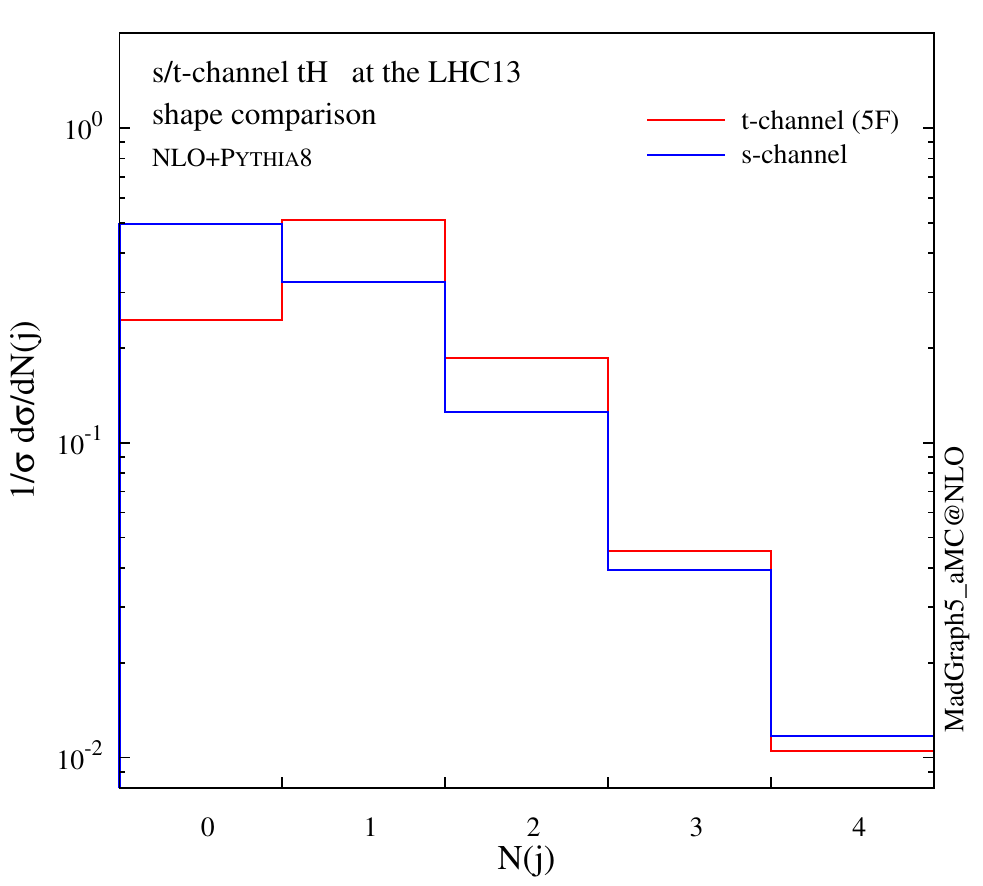}
 \includegraphics[width=0.325\textwidth]{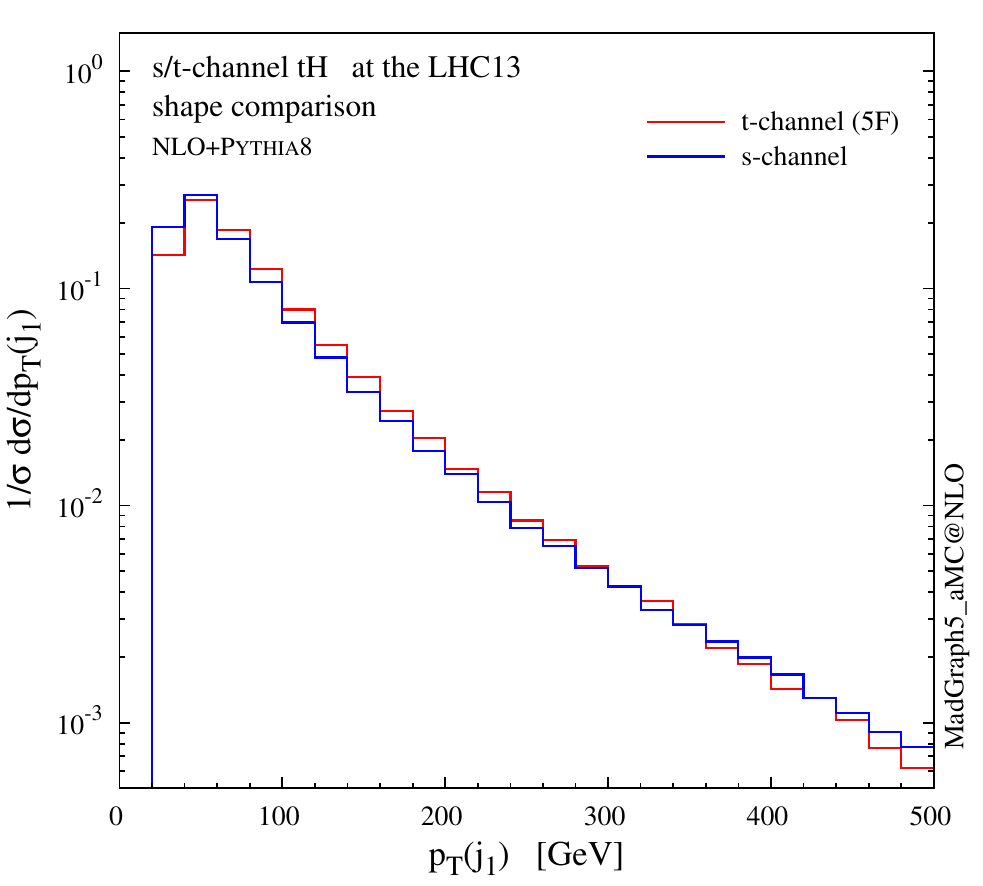}
 \includegraphics[width=0.325\textwidth]{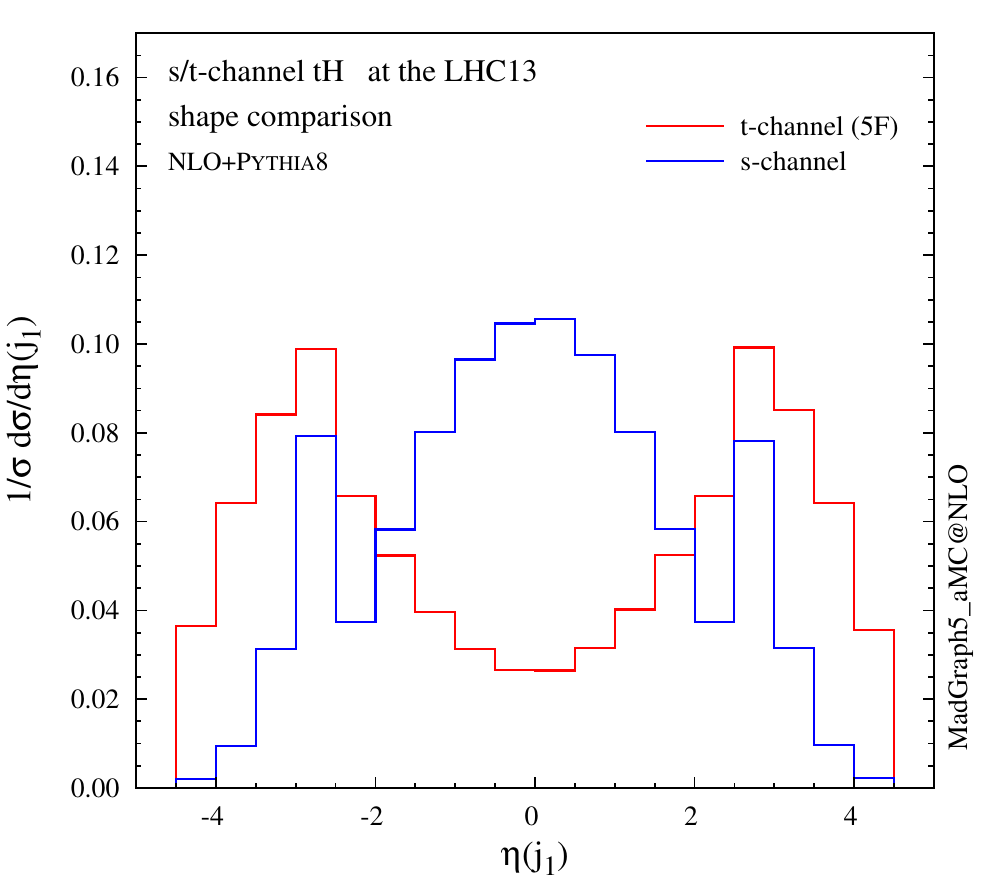}\\[1mm]
 \includegraphics[width=0.325\textwidth]{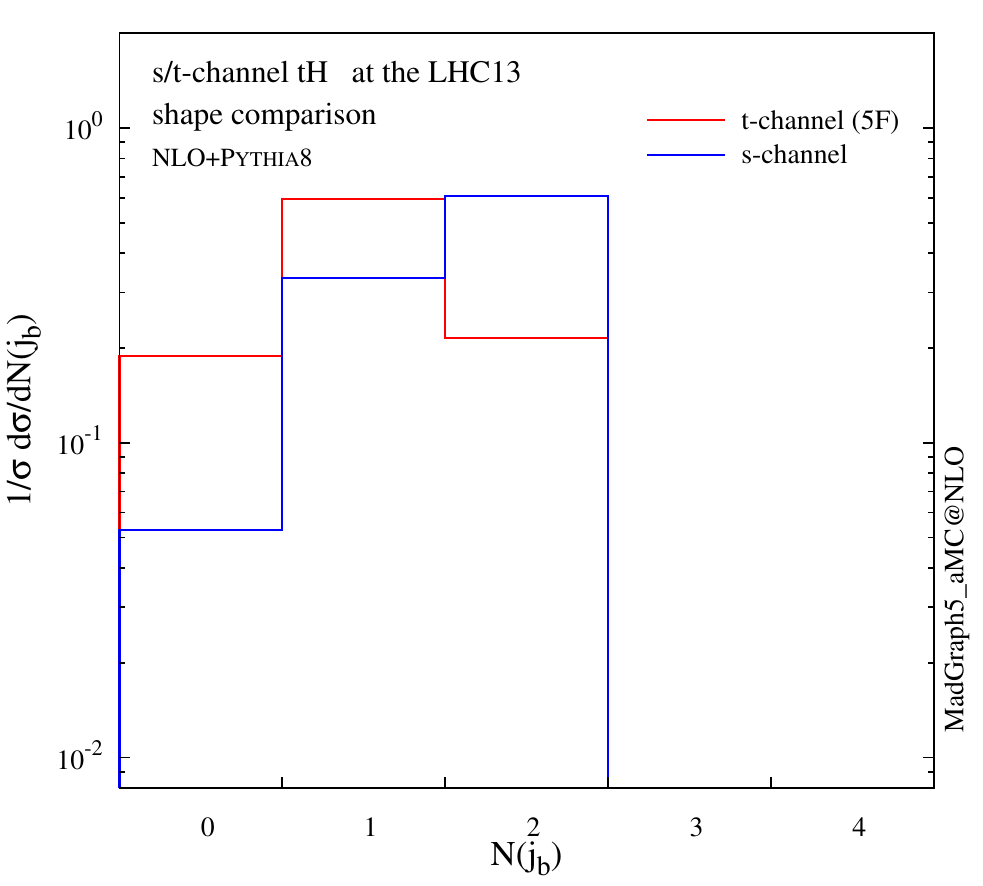}
 \includegraphics[width=0.325\textwidth]{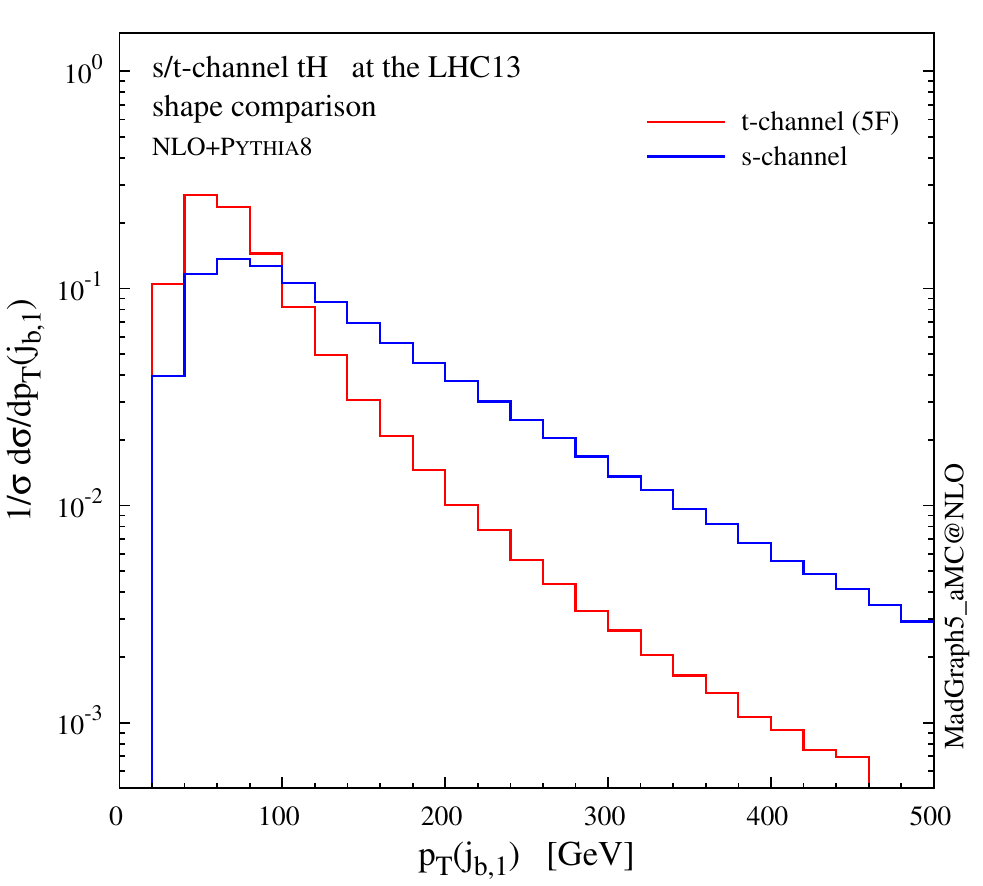}
 \includegraphics[width=0.325\textwidth]{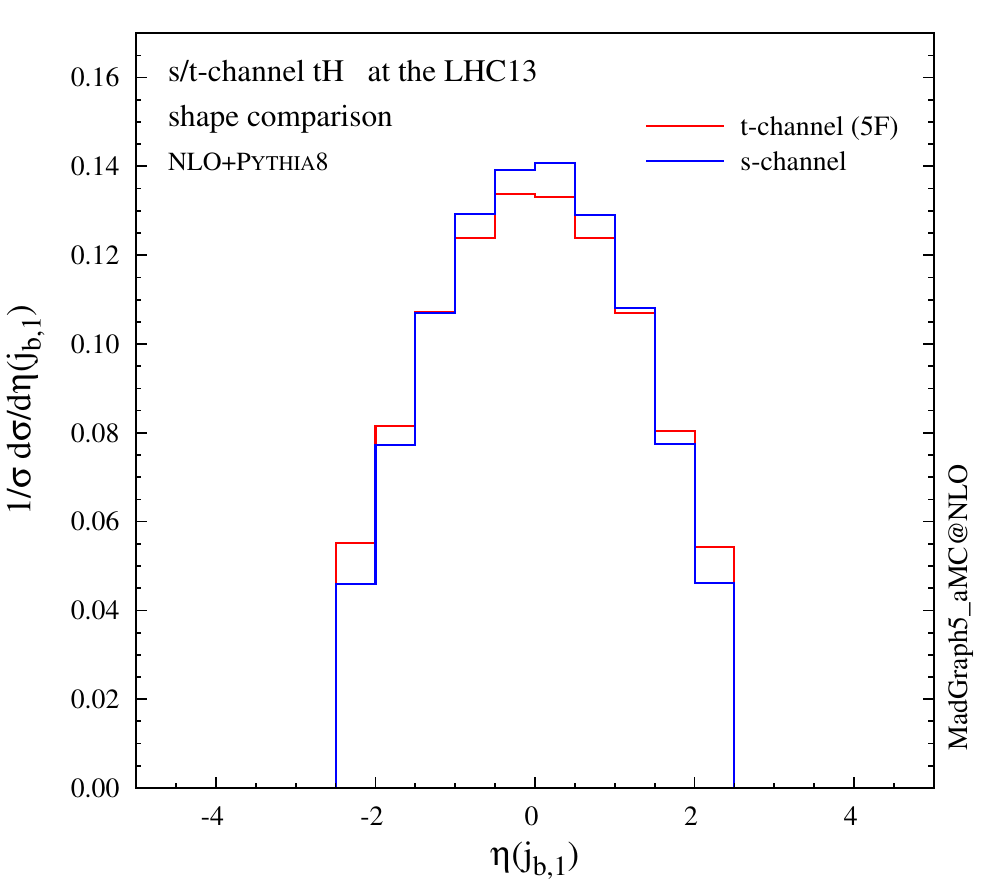}
 \caption{In the top row: shape comparison between $s$- and $t$-channel
 distributions for jet rates (left), $p_T$ (centre) and $\eta$ (right)
 spectra for the hardest jet at NLO+PS accuracy.
 In the bottom row: corresponding plots for $b$-tagged jets.}
\label{fig:STchannel_jets}
\end{figure*} 

At LO, $s$-channel production proceeds through $q \bar q$ annihilation
into a virtual $W$ boson, which can either emit a Higgs boson and then
split to a $tb$ final state, or first split to $tb$ with the subsequent
emission of a Higgs from the top quark. 
It turns out that in this case the interference between these two
diagrams is positive and its effect are much less relevant than in
$t$-channel production~\cite{Maltoni:2001hu}.  
At NLO, extra radiation can take place from either initial or final
state, with no interference between the two due to colour
conservation. 
For the same reason, no interference between the $s$-channel and
$t$-channel processes is present in the 5F scheme and the separation
between channels is still exact at NLO accuracy.  
In this production mode, bottom quarks are directly produced in the hard
scattering via electroweak interaction and appear only in the final
state.
Thus, at variance with the $t$-channel and $W$-associated production,
the flavour scheme is not a key source of uncertainties for $s$-channel
production. 

In the {\sc MadGraph5\_aMC@NLO} framework the code and the events 
for $s$-channel production at hadron colliders
can be automatically generated  by typing the following commands:
\begin{verbatim}
(> import model loop_sm)
 > generate p p > w+ > h t b~ [QCD]
 > add process p p > w- > h t~ b [QCD]
 > output
 > launch
\end{verbatim}

In table~\ref{tab:xsec_Schannel} we show the total cross section at
NLO. 
Reference values for the factorisation and renormalisation scales are
set to $\mu_0=H_T/2=\sum\,m_T/2\,$.
Being a pure EW process at LO, $s$-channel production exhibits very low
scale and $\alpha_s$ uncertainties up to NLO.
In the SM, the total rate amounts to about 3~fb, {\it i.e.} less than
$5\perc$ of the $t$-channel cross section. 

\begin{table}
\center
\begin{tabular}{r|llll}
 \hline
    \rule{0pt}{3ex}
    $s$-channel
  & $\sigma_{\rm NLO}$~{\small [fb]} 
  & $\delta^\perc_\mu$
  & $\delta^\perc_{\rm PDF}$
  & $\delta^\perc_{\alpha_s}$
    \\[0.7ex]
 \hline
     \rule{0pt}{3ex}
 $tH + \bar tH$ 
     & 2.812(3)
     & $^{+1.6}_{-1.2}$ 
     & $^{+1.4}_{-1.4}$
     & $^{+0.3}_{-0.5}$
    \\[0.7ex]
 \hline
\end{tabular}
 \caption{NLO total cross section for the processes
 $pp\to tH\bar b+\bar tHb$ via an $s$-channel $W$-boson exchange at the
 LHC ($\sqrt{s}=13$~TeV).
 {\sc NNPDF2.3} PDFs have been used. 
 The integration uncertainty in the last digit (in parentheses), the
 fractional scale dependence and the PDF and $\alpha_s$ uncertainties
 (in $\perc$) are also reported.}
\label{tab:xsec_Schannel}
\end{table}

In figs.~\ref{fig:STchannel_ht} and \ref{fig:STchannel_jets} we compare
the shape of some distributions between the $s$-channel and $t$-channel
production modes at NLO+PS accuracy.
We can see that most of the observables related to $s$-channel events
display a significantly different shape. 
Even though the total cross section in $s$-channel production is tiny
and deviations from a $t$-channel-only simulation would probably fall
inside the uncertainty band, the $s$-channel simulation can be included
with little extra computing cost when precision is needed (it is also
extremely fast at NLO).

%%%%%%%%%%%%%% Begin Section: HC %%%%%%%%%%%%%%%%%%%%%%%%%%%%%%%%%%%%%%%%% 
\section{Higgs characterisation}\label{sec:HC}

In this section we go beyond the SM and explore the sensitivity of Higgs-single-top associated production to a Higgs boson coupling to the top quark that does not conserve CP. Several phenomenological studies on anomalous Higgs coupling determination via Higgs-single-top associated production have appeared~\cite{Biswas:2012bd,Farina:2012xp,Yue:2014tya,Kobakhidze:2014gqa,Chang:2014rfa,Englert:2014pja,Ellis:2013yxa,Agrawal:2012ga}.
Current experimental constraints on the Higgs-boson couplings favour the SM, and
in particular for the top quark the magnitude is consistent with the SM expectations, even though
an opposite sign with respect to the SM one is not yet completely excluded~\cite{Khachatryan:2014jba,ATLAS-CONF-2014-009}.

Moreover, although the scenario of a pseudoscalar Higgs is 
disfavoured~\cite{ATLAS:2013mla,Khachatryan:2014kca},
no stringent constraint has been put on a CP-violating $Ht\bar t$ coupling.
In fact, even if current results are fully compatible with the SM hypothesis, 
some analyses on public LHC data seem to favour a non-zero  phase in the top quark Yukawa 
interaction~\cite{Djouadi:2013qya,Cheung:2013kla,Nishiwaki:2013cma,Boudjema:2015nda}.

In this work we consider the (simplified) case of a spin-0 particle with a general CP-violating Yukawa interaction with the top quark, which
couples both to scalar and pseudoscalar fermionic densities.  On the other hand, we assume the interaction 
with the $W$ bosons to be the SM one. We note that this assumption does
not correspond to a typical realisation of CP-violation in a
two-Higgs-doublet model (2HDM) where
the mass eigenstates are CP-mixed states and their coupling to the vector bosons is reduced. 
Our setup, however, corresponds to considering  the effective SM Lagrangian and to including the operator 
\begin{align}
 {\cal L}
 =\frac{c_t}{\Lambda^2} (\phi^\dagger \phi)\, Q_L \tilde \phi\, t_R +h.c.
\label{eq:dim6}
\end{align}
with $c_t$ complex. The implementation we use is based on  the effective
field theory framework presented in
refs.~\cite{Artoisenet:2013puc,Maltoni:2013sma,Demartin:2014fia} and
employs the {\sc HC\_NLO\_X0} model\cite{FR-HC:Online}.%
\footnote{For the code and event generation, one can simply issue the
command `{\tt import model HC\_NLO\_X0}' and replace 
`{\tt h}' by `{\tt x0}' in {\sc MadGraph5\_aMC@NLO}.}   
The effective Lagrangian for the Higgs-top quark interaction \eqref{eq:dim6} below the EWSB scale leads to   (see eq.~(2.2) in ref.~\cite{Artoisenet:2013puc})
\begin{align}
 {\cal L}_0^t 
   = -\bar\psi_t\big(
         c_{\alpha}\kappa_{\sss Htt}g_{\sss Htt} 
       +i s_{\alpha}\kappa_{\sss Att}g_{\sss Att}\, \gamma_5 \big)
      \psi_t\, X_0 \,,
\label{eq:0ff}
\end{align}
where $X_0$ labels a generic spin-0 particle with CP-violating couplings, 
$c_{\alpha}\equiv\cos\alpha$ and $s_{\alpha}\equiv\sin\alpha$ are related to the 
CP-mixing phase $\alpha$, 
$\kappa_{\sss Htt,Att}$ are real dimensionless rescaling parameters, and
$g_{\sss Htt}=g_{\sss Att}=m_t/v\,(=y_t/\sqrt{2})$, with $v\sim 246$~GeV.  
While redundant (only two independent real quantities are needed to parametrise 
the most general CP-violating interaction with the top quark at dimension four), 
this parametrisation has the practical advantage of 
easily interpolating  between the CP-even ($c_{\alpha}=1,s_{\alpha}=0$) and
CP-odd  ($c_{\alpha}=0,s_{\alpha}=1$) couplings, 
as well as to easily recover the SM case by setting $c_{\alpha}=1 \,,\, \kappa_{\sss Htt}=1 \,$.

\begin{figure}
\center 
 \includegraphics[width=0.495\textwidth]{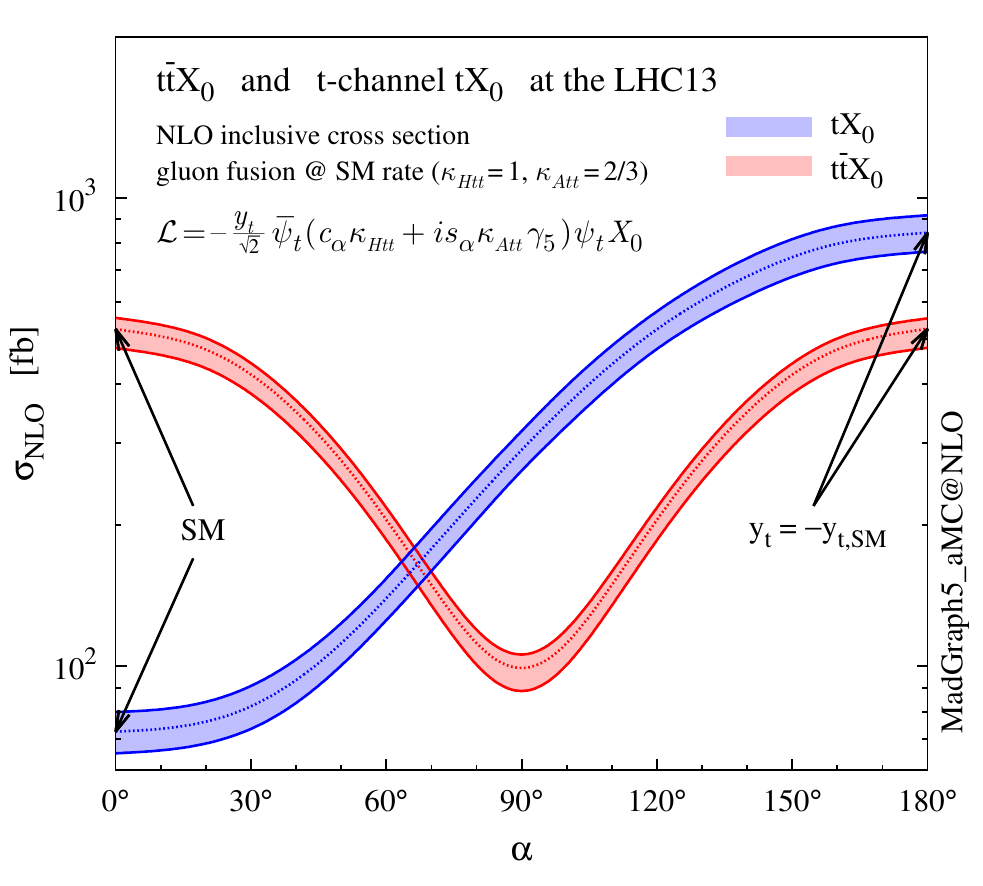}
 \caption{NLO cross sections (with scale uncertainties) for $t\bar tX_0$
 and $t$-channel $tX_0$ productions at the 13-TeV LHC as a function of
 the CP-mixing angle $\alpha$, where $\kappa_{\sss Htt}$ and 
 $\kappa_{\sss Att}$ are set to reproduce the SM GF cross section for
 every value of $\alpha$.}   
\label{fig:hc_xsect}
\end{figure} 

\begin{figure*}
\center 
 \includegraphics[width=0.325\textwidth]{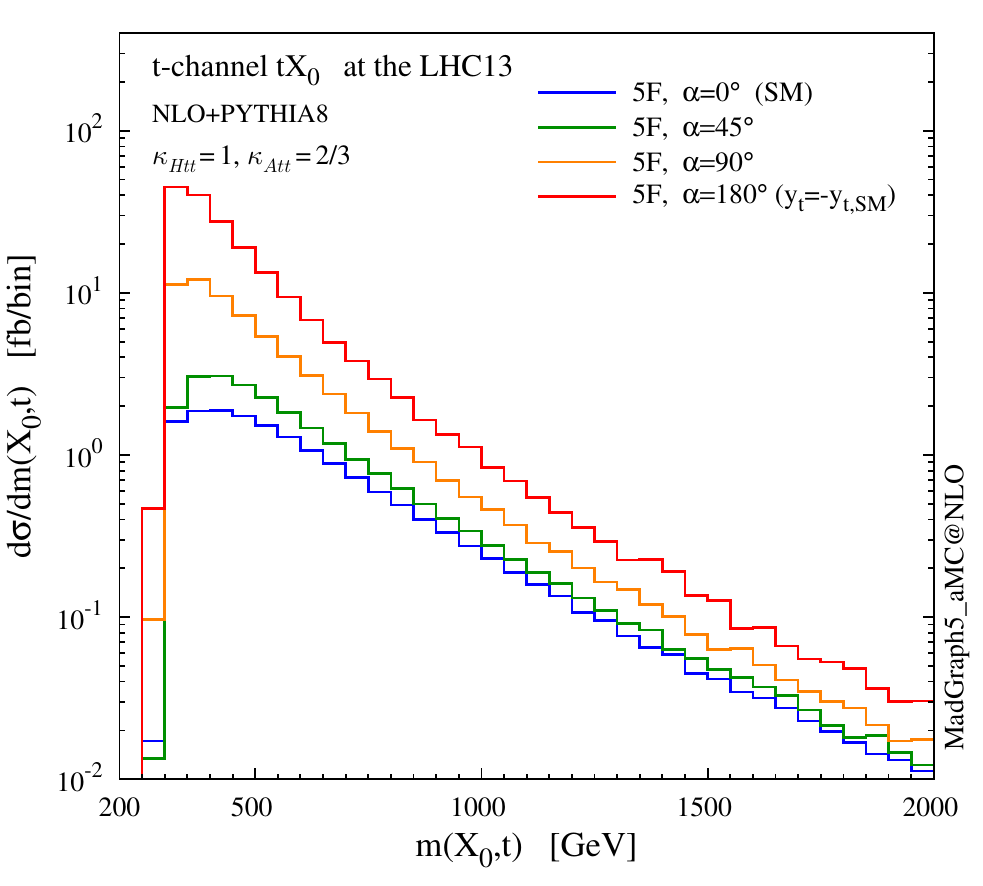}
 \includegraphics[width=0.325\textwidth]{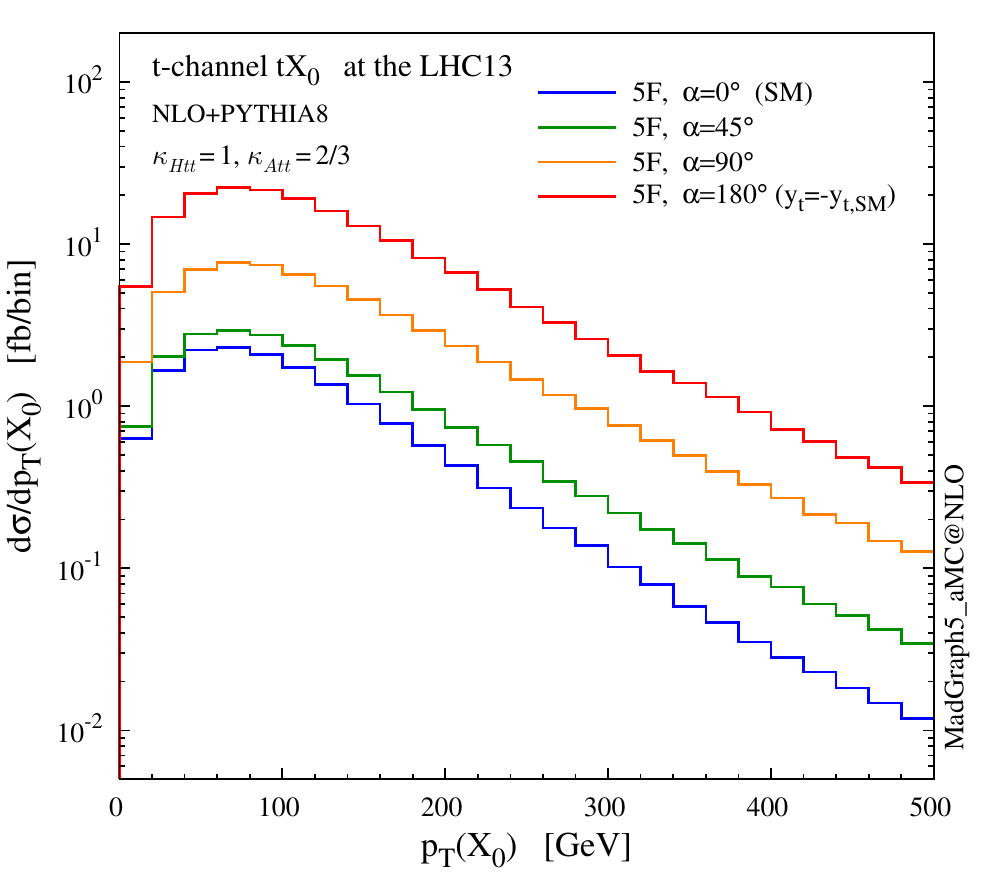}
 \includegraphics[width=0.325\textwidth]{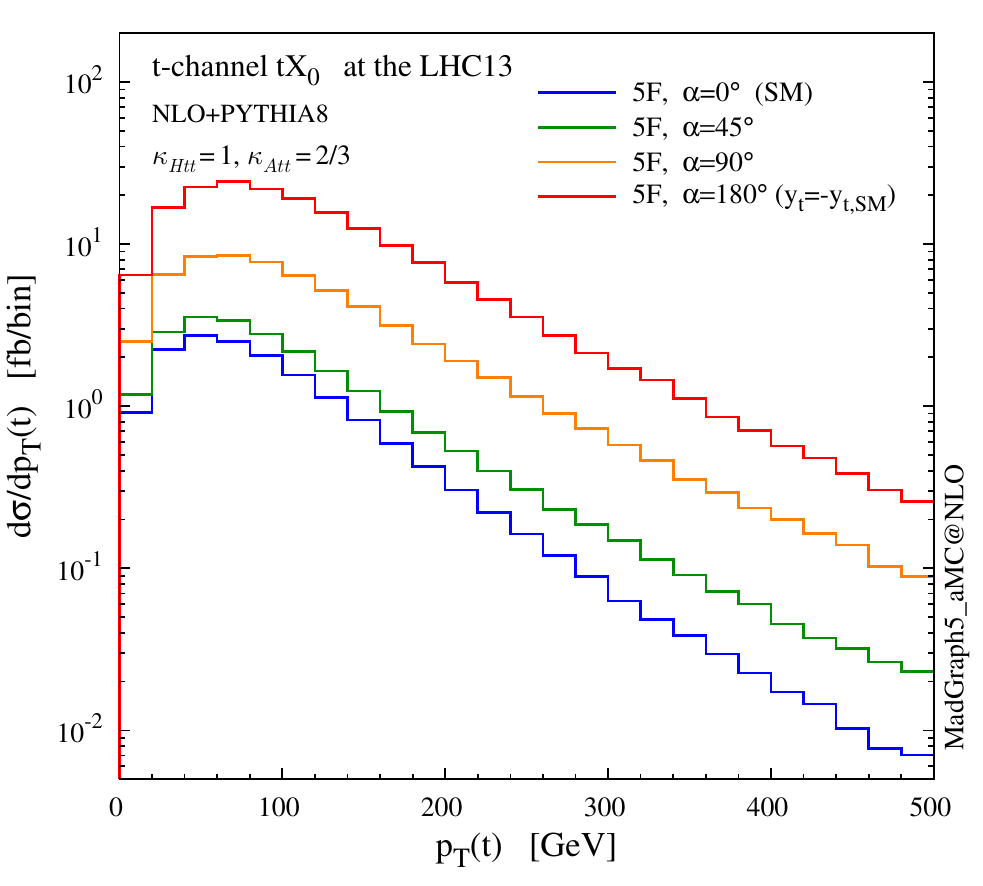}
 \caption{Differential distributions for the Higgs boson and the top
 quark at NLO+PS accuracy in $t$-channel $tH$ associated 
 production at the 13-TeV LHC, with different values of the CP-mixing
 angles, where $\kappa_{\sss Htt}$ and 
 $\kappa_{\sss Att}$ are set in eq.~\eqref{eq:GFkappas} to reproduce the SM GF cross section for
 every value of $\alpha$.} 
\label{fig:HC_distrib1}
\end{figure*} 

\begin{figure*}
\center 
 \includegraphics[width=0.325\textwidth]{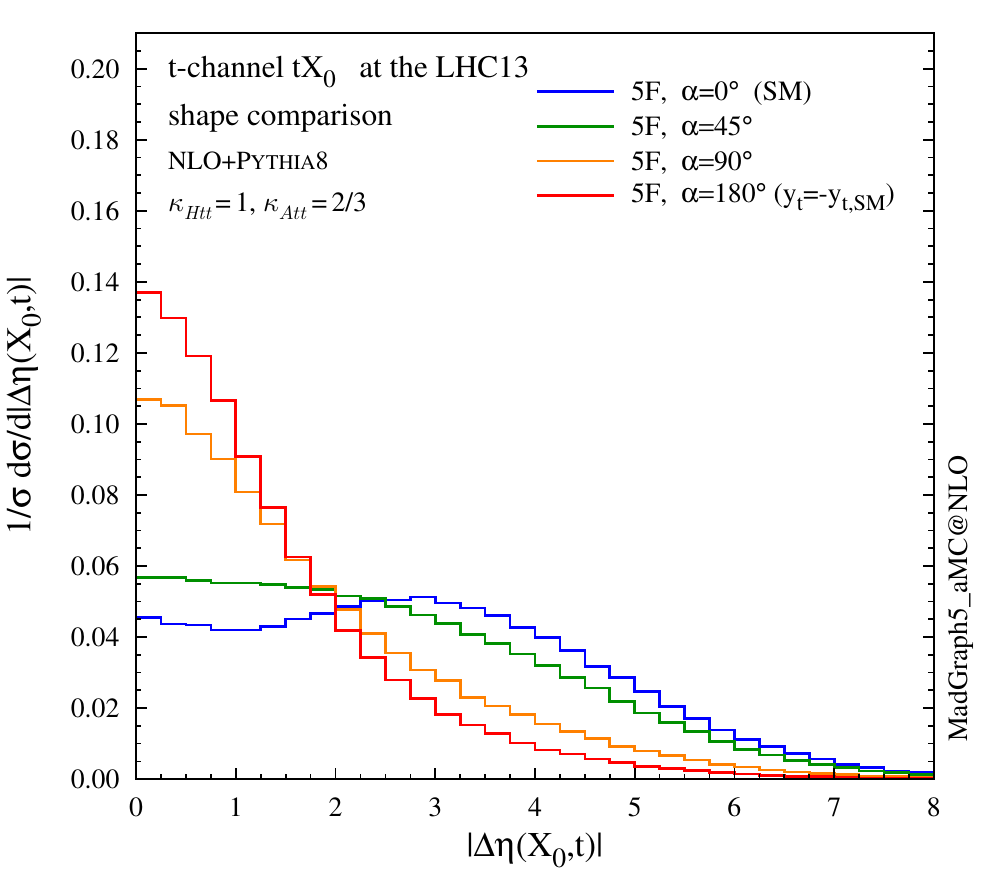}
 \includegraphics[width=0.325\textwidth]{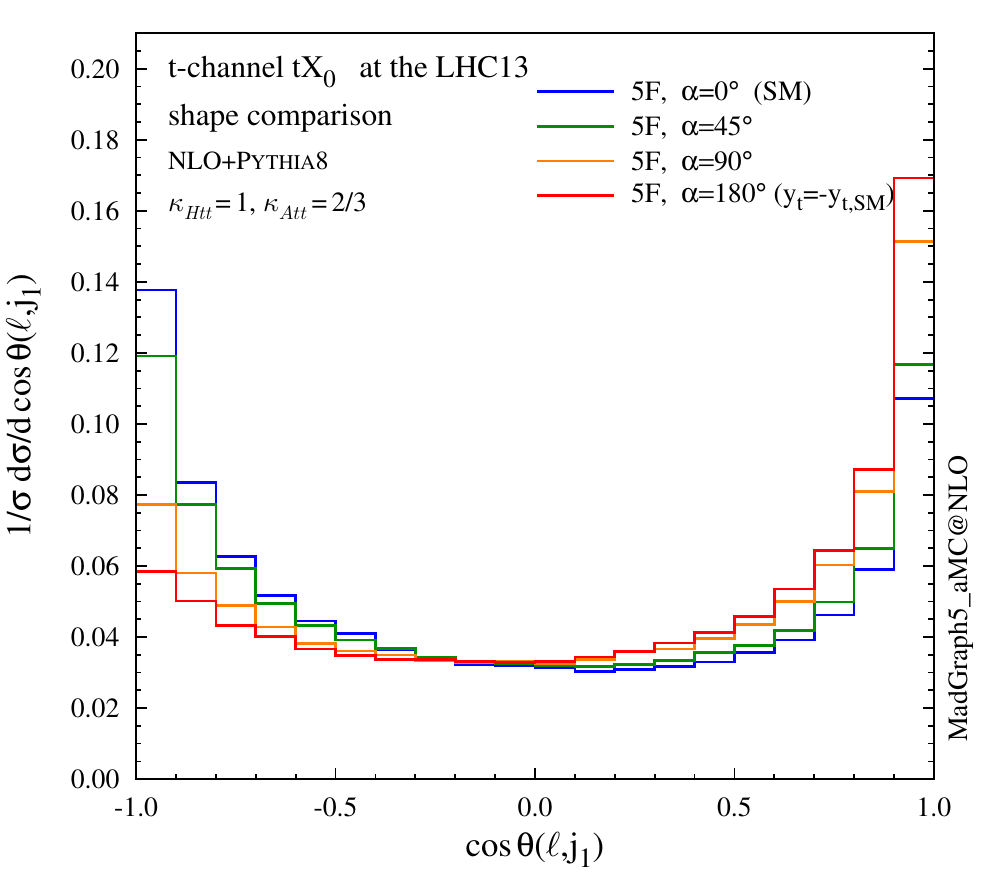}
 \caption{Shape comparison among different values of the CP-mixing
 angles, where $\kappa_{\sss Htt}$ and 
 $\kappa_{\sss Att}$ are set in eq.~\eqref{eq:GFkappas} to reproduce the SM GF cross section for
 every value of $\alpha$. 
 Pseudorapidity separation between the Higgs and the top quark (left)
 and opening angle between the hardest jet and the lepton from the
 top quark in the lab frame (right).}  
\label{fig:HC_distrib2}
\end{figure*} 

The nature of the top quark Yukawa coupling directly affects  the loop-induced 
Higgs coupling to gluons (together with an effect on the couplings to $\gamma\gamma$ and $Z\gamma$,
which are also modified but not considered here)
\begin{align} 
 {\cal L}_0^{g} &=
  -\frac{1}{4}\big(c_{\alpha}\kappa_{\sss Hgg}g_{\sss Hgg} \,
  G_{\mu\nu}^aG^{a,\mu\nu} \nn\\
   &\hspace*{1cm}+s_{\alpha}\kappa_{\sss Agg}g_{\sss Agg}\,G_{\mu\nu}^a\widetilde G^{a,\mu\nu} \big) 
  X_0\,,
\label{eq:L_Hglu}
\end{align}
where $ g_{\sss Hgg} = -\alpha_s/(3\pi v) $ and $ g_{\sss Agg} = \alpha_s/(2\pi v) $.
In the parametrisation given above, the strength of the coupling between Higgs
and gluons can be rescaled independently of the top quark Yukawa coupling. Assuming that the 
the top quark dominates the gluon-fusion (GF) process at the LHC energies, then $\kappa_{\sss Hgg} \to \kappa_{\sss Htt} \,$,
$\kappa_{\sss Agg} \to \kappa_{\sss Att} \,$. 
In so doing, the ratio between the actual cross section for GF
at NLO QCD and the corresponding SM prediction can be written as
\begin{align}
\frac{\sigma_{\rm NLO}^{gg \to X_0} }{\sigma_{\rm NLO,SM}^{gg \to H}} \, =\, 
 c^2_\alpha \, \kappa^2_{\sss Htt} \,+\, 
 s^2_\alpha \Big( \kappa_{\sss Att} \, \frac{g_{\sss Agg}}{g_{\sss Hgg}} \Big)^2 \,,
\label{eq:GFrate}
\end{align}
because there is no interference between the scalar and pseudoscalar components
in the amplitudes for Higgs plus up to three external partons, see e.g.,~\cite{Demartin:2014fia}.
In particular, if the rescaling parameters are set to 
\begin{align}
 \kappa_{\sss Htt} = 1 \,, \qquad \kappa_{\sss Att} 
 = |\, g_{\sss Hgg}/ g_{\sss Agg} \,| = 2/3 \,,
\label{eq:GFkappas}
\end{align}
the SM GF cross section is reproduced for every value of the CP-mixing
phase $\alpha$. 
Given that current measurements are compatible with the expected SM
GF production rate, one can consider the simplified scenario where the
condition in eq.~\eqref{eq:GFkappas} is imposed and the CP-mixing phase
$\alpha$ is basically left unconstrained by current data.

Figure~\ref{fig:hc_xsect} shows the total cross section for $t$-channel 
$tX_0$ production as a function of the CP-mixing angle $\alpha$. 
We also show the $t\bar tX_0$ cross section, which is not only another
process sensitive to the modifications of the top quark Yukawa coupling
in eq.~\eqref{eq:0ff}, but also a background to $t$-channel production.
The uncertainty band represents the envelope defined in
sect.~\ref{sec:totrates}, {\it i.e.} the combined scale and
flavour-scheme dependence. 
The $t\bar tX_0$ uncertainty band represents the scale dependence only,
when the scale is varied by a factor two around
$\mu_0=\sqrt[3]{m_T(t)\,m_T(\bar t)\,m_T(X_0)}$~\cite{Demartin:2014fia}.  

The first important observation is that while the GF and $t\bar tH$
cross sections are degenerate under $y_t\to-y_t$ (depending
quadratically from the top quark Yukawa coupling), in $t$-channel
production this degeneracy is clearly lifted by the interference between
diagrams where the Higgs couples to the top quark and to the $W$ boson.
In~\cite{Biswas:2012bd,Farina:2012xp} it was shown that the $t$-channel
cross section is enhanced by more than one order of magnitude when the
strength of the top Yukawa coupling is changed in sign with respect to
the SM value. 
Here we can see how the same enhancement can take place also in the
presence a continuous rotation in the scalar-pseudoscalar plane.
While not affecting GF (by construction), such a rotation has an impact
also on the $t\bar tX_0$ rate, which is in general lower for a
pseudoscalar or CP-mixed state~\cite{Demartin:2014fia}. 
$t$-channel production lifts another degeneracy present in GF and 
$t\bar t X_0$, namely $\alpha\to\pi-\alpha\,$.
Given the partial compensation between the $t$-channel and
$t\bar tX_0$ cross sections at different values of $\alpha$, an analysis
which could well separate between the two production mechanisms would be
needed to put stringent constraints on a CP-violating Higgs coupling to
the top quark. 

We remind that the enhancement of the $t$-channel cross section
takes place mostly at threshold, as one can clearly see in the left plot
of fig.~\ref{fig:HC_distrib1}.
This means that one should not be concerned by violations of
perturbative unitarity at the LHC, as they do not appear for partonic
centre-of-mass energies lower than $\sim 10$~TeV~\cite{Farina:2012xp}. 
In fig.~\ref{fig:HC_distrib1} we also show the transverse momentum
distributions for the Higgs and the top quark.
The distributions are well behaved in this case too, not displaying any
strong trend in their high-$p_T$ tails, {\it i.e} anything that could
suggest a unitarity violating behaviour. 

Finally, in fig.~\ref{fig:HC_distrib2} we plot the pseudorapidity
separation between the Higgs and the top quark (left) and the opening
angle between the hardest jet and the lepton from the top quark in the
lab frame (right), showing that these variables have a discriminating
power on $\alpha$. 
For this last observable, the lepton is required to satisfy the
following selection criteria 
\begin{align}
\label{eq:lep_definition}
 & p_T(\ell)>20~{\rm GeV}\,, \quad |\eta(\ell)|<2.5\,.
\end{align}
%

%%%%%%%%%%%%%% Begin Section: Summary %%%%%%%%%%%%%%%%%%%%%%%%%%%%%%%%%%%%%%%%% 
\section{Summary}\label{sec:summary}

In this work we have studied the production of a Higgs boson in association with a single top quark at
the LHC. Our aim has been to carefully consider the effects of  NLO corrections in QCD  on total cross sections and differential distributions for $t$- and $s$-channel production. We have scrutinised a wide range of theoretical 
systematic uncertainties and in particular those arising from the choice of the heavy-quark scheme, 4-flavour or 5-flavour.
We have found that at the level of total cross sections a comfortable consistency between the two schemes
exists when physically motivated choices for the renormalisation and factorisation scales are made, with similar
resulting uncertainties. For differential distributions, on the other hand, the situation is slightly more involved. 
While sizeable differences between the two schemes arise at LO, they are
considerably milder at NLO and NLO+PS, in line with expectations. In this case, we have shown that the 4F and 5F schemes provide  fully consistent and similarly precise predictions for distributions such as those of the Higgs boson, the top quark, and the forward jet.  On the other hand, the 4-flavour scheme is in general able to provide accurate predictions for a wider set of observables, including  those of the spectator $b$-quark and extra jets. 
In addition to $t$-channel production in the SM, we have also briefly presented the results for the 
subdominant $s$-channel production, highlighting the differences in the most important distributions
with respect to the corresponding ones of $t$-channel production. 
Finally, we have provided results (total cross sections as well as a few
representative distributions) for the case where an explicit CP
violation is present in the coupling between the top quark and the Higgs
boson, making it clear that in this case Higgs associated production
with a single top could provide complementary and very valuable
information to that of $t \bar t H$ production. We conclude by stressing
that all results presented here have been obtained by employing the publicly
available {\sc MadGraph5\_aMC@NLO} framework and therefore they can be
easily reproduced  (and possibly extended) by generating  the
corresponding event samples to be used in fully-fledged experimental
analyses.

%AAAAAAAAAAAAAAAAAAAAAAAAA
\section*{Acknowledgments}

We thank the LHC HXSWG and in particular the members of the $ttH/tHj$ task force 
for giving us the motivation to pursue this study and for many stimulating discussions. 
We are grateful to Rikkert Frederix and Stefano Frixione for discussions and to Paolo Torrielli for his valuable help. 
This work has been performed in the framework of the ERC grant 291377
``LHCtheory: Theoretical predictions and analyses of LHC physics:
advancing the precision frontier''  and of the FP7 Marie Curie Initial Training 
Network MCnetITN (PITN-GA-2012-315877). It is also supported in part by the 
Belgian Federal Science Policy Office through the Interuniversity Attraction Pole P7/37. 
The work of FD and FM is supported by the IISN ``MadGraph'' convention
4.4511.10 and the IISN ``Fundamental interactions'' convention 4.4517.08.
KM is supported in part by the Strategic Research Program ``High Energy
Physics'' and the Research Council of the Vrije Universiteit Brussel.  
The work of MZ is supported by the ERC grant ``Higgs@LHC'' and partially by the ILP LABEX (ANR-10-LABX-63), 
in turn supported by  French state funds managed by the ANR within the ``Investissements d'Avenir''
programme under reference ANR-11-IDEX-0004-02.

% APPENDIXAPPENDIXAPPENDIXAPPENDIXAPPENDIXAPPENDIXAPPENDIXAPPENDIXAPPENDIX  
%\appendix

%RRRRRRRRRRRRRRRRRRRRRRRRRRRRRRRRRRRRRRRRRRRRRRRRRRRRRRRRRRRRR
%\bibliography{library}
%\bibliographystyle{spphys}
%\bibliographystyle{JHEP}
\providecommand{\href}[2]{#2}\begingroup\raggedright\endgroup

\end{document}